\documentclass[]{aa}  

\usepackage{graphicx}
\usepackage{txfonts}
\usepackage{graphicx}
\usepackage{txfonts}
\usepackage{amsmath}
\usepackage{adjustbox}
\usepackage{hyperref}
\usepackage{gensymb}
\usepackage{placeins}
\usepackage{comment}
\usepackage{lipsum}
\usepackage{subcaption}
\hypersetup{colorlinks=true,allcolors=[rgb]{0,0,0.5}}
\newcommand*{\plimsoll}{{\ensuremath{-\kern-5pt{\circ}\kern-5pt-}}}

\makeatletter
\renewcommand*\aa@pageof{, page \thepage{} of \pageref*{LastPage}}
\makeatother
\begin{document}

   \title{Vanadium oxide clusters in substellar atmospheres}

   \subtitle{A quantum chemical study}

   \author{H. Lecoq-Molinos\inst{1, 2, 3}
          \and
          D. Gobrecht \inst{4}
          \and
          J.P. Sindel \inst{1,2,5,6}
          \and
          Ch. Helling \inst{1,3}
          \and
          L. Decin \inst{2}}

   \institute{   Space Research Institute, Austrian Academy of Sciences, Schmiedlstrasse 6, A-8042 Graz, Austria\\
              \email{helena.lecoq@oeaw.ac.at}
         \and
       Institute of Astronomy, KU Leuven, Celestijnenlaan 200D, 3001 Leuven, Belgium
         \and
             TU Graz, Fakult\"at f\"ur Mathematik, Physik und Geod\"asie, Petersgasse 16, A-8010 Graz, Austria
         \and
       Department of Chemistry and Molecular Biology, University of Gothenburg, Sweden
         \and
      Centre for Exoplanet Science, University of St Andrews, North Haugh, St Andrews, KY169SS, UK
       \and
       SUPA, School of Physics \& Astronomy, University of St Andrews, North Haugh, St Andrews, KY169SS, UK
             }
   \date{Received....; accepted .....}

 
  \abstract
  {As a refractory material vanadia (solid V$_2$O$_5$) is a likely condensate in the atmospheres of substellar objects such as exoplanets and brown dwarfs. However, the nature of the nanometer-sized vanadium oxide clusters, that partake in the nucleation process, is not well understood.} 
   {We aim to understand the formation of cloud condensation nuclei in oxygen-rich substellar atmospheres by calculating fundamental properties of the energetically most favorable vanadium oxide molecules and clusters and, to investigate how they contribute to the formation of condensation seeds. }
   {A hierarchical optimization approach is applied in order to find the most favourable structures for clusters of (VO)\textsubscript{N} and (VO$_2$)\textsubscript{N} for N=1-10, and (V$_2$O$_5$)\textsubscript{N} for N=1-4 and to calculate their thermodynamical potentials. The candidate geometries are initially optimized applying classical interatomic potentials and then refined at the B3LYP/cc-pVTZ level of theory to obtain accurate zero-point energies and thermochemical quantities. 

   }
   { We present previously unreported vanadium oxide cluster structures as lowest-energy isomers. Moreover, we report revised cluster energies and their thermochemical properties. Chemical equilibrium calculations are used to asses the impact of the updated and newly derived thermodynamic potentials on the gas-phase abundances of vanadium-bearing species. In chemical equilibrium, larger clusters from different stoichiometric families are found to be the most abundant vanadium-bearing species for temperatures below $\sim$1000 K, while molecular VO is the most abundant between $\sim$1000 K and $\sim$2000 K. We determine the nucleation rates of each stoichiometric family for a given (T$_{gas}$, p$_{gas}$) profile of a brown dwarf using classical and non-classical nucleation theory.  

   }
   {Small differences in the revised Gibbs free energies of the clusters have a large impact on the abundances of vanadium bearing species in chemical equilibrium at temperatures below $\sim$1000 K, which subsequently has an impact on the nucleation rates of each stoichiometric family. We find that with the revised and more accurate cluster data non-classical nucleation rates are up to 15 orders of magnitude higher than classical nucleation rates.}

   \maketitle
%

\section{Introduction}

\label{sec:Introduction}
   
The presence of dust and clouds throughout the Universe has a strong impact on the physics and chemistry of various astrophysical environments \citep{henning_astromineralogy_2010}. Dust is known to form in the outflows of asymptotic giant branch (AGB) stars \citep{sedlmayr_molecules_1994, ferrarotti_mineral_2002, gail_seed_2013, gobrecht_dust_2016, khouri_study_2016, decin_study_2017}, supernovae ejecta \citep{patzer_dust_1998, sarangi_dust_2018, zhang_sn_2021} and protoplanetary disks \citep{liu_tale_2020}, whereas clouds are present in the atmospheres of brown dwarfs \citep{ackerman_precipitating_2001, woitke_dust_2004, helling_dust_2006, suarez_ultracool_2023} and exoplanets \citep{charbonneau_detection_2002, kreidberg_clouds_2014, lee_dust_2014,sing_hst_2016, gao_aerosols_2021, helling_exoplanet_2023, samra_clouds_2023, feinstein_early_2023}.  
However, the formation processes of both clouds and dust grains remain unclear. In particular, the initial stages of the process, where gas-phase molecules recombine to form clusters, are not well understood.\\
In substellar atmospheres, these cluster particles provide the cloud condensation nuclei (CCN) necessary to onset cloud formation \citep{hudson_cloud_1993, lee_dust_2018, helling_exoplanet_2019, sindel_revisiting_2022}. On rocky planets CCN are provided by processes that take place at the surface of the planet, i.e volcanic activity or sandstorms. In gaseous exoplanets and brown dwarfs, the CCN formation process entails a transition from the gas phase to the solid phase. From a microscopic perspective, the transition is initiated by reactions between molecules, atoms or ions that lead to the formation of increasingly more complex species, i.e. nanometer-sized clusters.
The clusters will then continue to interact and grow until macroscopic solid particles are formed \citep{gail_dust_1984, gail_growth_2013, lee_dust_2018, boulangier_developing_2019} and provide a surface for other gas-phase molecules to condense on and form a dust grain or a cloud particle.
Clusters can therefore be considered as a bridge between the gas and the solid phase, filling the gap between small molecules and bulk materials \citep{johnston_atomic_2002}. The formation of clusters is commonly referred to as the \emph{nucleation} process.\\ 
There are two main approaches to the nucleation process: top-down \citep{schoiswohl_vanadium_2006, gail_seed_2013, de_jesus_defining_2018} and bottom up \citep{bromley_under_2016, gobrecht_bottom-up_2022, 
sindel_revisiting_2022}. In the top-down approach the macroscopic bulk properties are taken as a starting point and they are extrapolated to nanoscale structures. However, the size reduction 
from macroscopic to nanometer length scales can lead to drastic changes in the material properties (e.g. atomic ordering in the lattice, stability of the material, melting temperatures)
\citep{vines_size_2017} since quantum and surface effects on the small clusters lead to non-crystalline structures, whose characteristics (geometry, energy, coordination, spectra) deviate significantly from the crystalline bulk material. In a bottom-up approach the opposite direction is followed: we start with a 
single molecule or monomer and gradually increase in size towards the bulk limit. Clusters need to reach a certain size before the regular, periodic atomic ordering characteristic of the crystalline bulk material is thermodynamically favoured. Hence, the first steps of the nucleation process are more accurately described by a bottom-up approach, starting with the smallest molecules and using progressively larger clusters as  building blocks.\\ 
Even though there are different theories to describe the nucleation process, such as classical (CNT) \citep{gail_primary_1986}, modified (MCNT) \citep{gail_dust_1984, draine_timedependent_2008, helling_dust_2006, gail_growth_2013} and non-classical \citep{gail_growth_2013, helling_modelling_2013, lee_dust_2014} nucleation theory, as well as kinetic nucleation networks \citep{patzer_dust_1998, gobrecht_dust_2016, boulangier_developing_2019, gobrecht_bottom-up_2022}, they all require thermochemical data of the nucleating species, in particular of their respective clusters. 
Experimental data is commonly available for the condensed species as well as simple gas-phase molecules but it becomes sparse for the nanocluster size regime. Nucleation has been studied experimentally \citep{li_understanding_2021}. However, such experiments are limited in the range of pressures and temperatures that do not cover the conditions in substellar atmospheres. Therefore, these experiments cannot provide information regarding which materials are more likely to nucleate under substellar conditions.
Metal oxide cluster ions can be synthesized in cluster beam experiments (see e.g., \citet{asmis_mass-selective_2007, fielicke_stability_2002,marinoso_guiu_cluster_2022}), but we aim to study neutral clusters.
Quantum mechanical calculations provide a way to bridge the gap between the gas and the solid phase data. Global optimizations can be performed to determine the most favourable geometry for each cluster size, which is then assumed to take part in the nucleation process. Several studies have applied a quantum-mechanical bottom-up approach to the formation of CCN for a diverse range of chemical species, including Ti-bearing species such as TiO$_2$ \citep{sindel_revisiting_2022, jeong_electronic_2000}, alumina \citep{gobrecht_bottom-up_2022}, SiO \citep{bromley_under_2016} and Fe-bearing species \citep{chang_small_2013}. Vanadium oxide compounds have been a common subject of study in materials science \citep{asmis_gas_2004, asmis_mass-selective_2007, janssens_isomorphous_2006, archambault_density-functional_2021, vyboishchikov_gas-phase_2000} due to their various technological applications \citep{weckhuysen_chemistry_2003, whittingham_lithium_2004, krusin-elbaum_room-temperature_2004} but their consideration as nucleation species in substellar atmospheres is unprecedented.\\ 
The vanadium monoxide molecule has been observed in the atmospheres of red giant stars \citep{alvarez_near-infrared_1998} and exoplanets \citep{evans_detection_2016, pelletier_vanadium_2023}, whereas solid vanadium oxide is found in pristine meteoric material \citep{rehder_possible_2011}. As refractory materials, vanadium oxides are alternative candidates to classical seed particle species that include oxides of silicon \citep{gail_seed_2013, bromley_under_2016}, titanium \citep{sindel_revisiting_2022, jeong_electronic_2000}, and aluminum \citep{gobrecht_bottom-up_2022, patzer_density_2005}. Perhaps it is due to the comparatively low solar vanadium abundance that, to our knowledge, no systematic study on the nucleation of vanadium oxides has been carried out before. Vanadium oxides exhibit relatively low vapor pressures and correspondingly high boiling temperatures (see e.g., \citet{haynes_crc_2016}), which is also reflected in the high VO bond energy ($\sim$ 630 kJ mol$^{-1}$ \citep{balducci_thermochemical_1983, merriles_bond_2020}). Furthermore, molecular VO is observed only in the hottest substellar atmospheres indicating that VO is incorporated into cloud particles in environments with lower temperatures. Therefore, vanadium oxides are promising nucleation (i.e., condensation seed) candidates at temperatures around $\sim$ 1000 K, where dust nucleation occurs in astrophysical environments.\\
In this study we aim to assess the viability of vanadium oxides as nucleation seeds in substellar atmospheres. In Sect. \ref{sec:Methods} we go over the computational chemistry methods applied to obtain the most stable cluster for several vanadium oxide compounds ( (V\textsubscript{x}O\textsubscript{y})\textsubscript{N}, N=1-10, x=1,2, y=1,2,5). The thermochemical characteristics of the lowest-energy isomers and the impact of the updated cluster data in chemical equilibrium calculations are presented in Sect. \ref{sect:results}. In Sect. \ref{sect:astro_rel} we calculated the nucleation rates for a brown dwarf model atmosphere and we present the vibrational spectra for each isomer. Finally, we summarize our results and further work in Sect. \ref{sect:conclusions}.

\section{Methods}
\label{sec:Methods}
The subjects of this paper are vanadium oxide clusters with different stoichiometries, i.e. (V\textsubscript{x}O\textsubscript{y})\textsubscript{N}, N=1-10, x=1,2, y=1,2,5. The selection of the clusters is based on our computational capabilities and the stability of the stoichiometric families in the gas-phase and as a solid. We investigate their geometries, binding energies, thermochemical properties and equilibrium abundances.\\
In order to obtain the data we conducted a search of the most favorable isomers for each size and stoichiometry. The most favorable isomer for each cluster size represents the lowest minimum on the potential energy surface (PES) and we will refer to it as the global minimum (GM) candidate. To find the global minima candidate we use a large number of initial geometries of each considered cluster size and stoichiometry until we find the lowest energy configuration. The number of possible structural isomers increases rapidly with the clusters size, i.e. the number of atoms in the cluster. Therefore, an exploration of the entire PES at the Density Functional Theory (DFT) level of theory is computationally very demanding for clusters with tens of atoms. To minimize the computational effort, we employ a hierarchical global optimization technique using an interatomic potential as first step in our (V\textsubscript{x}O\textsubscript{y})\textsubscript{N} cluster study (see Sect. \ref{sec:methods_FFM}). This procedure reduces the number of possible structural configurations significantly and generates low-energy candidate isomers. The lowest-energy candidate structures are subsequently refined by optimizations at the DFT level (see Sect. \ref{sec:methods_DFT}).

\subsection{Optimization with force fields}
\label{sec:methods_FFM}
We have first performed a global optimization search for low-energy isomers using the interatomic Buckingham-Coulomb pair potential. The optimizations were computed with the General Utility Lattice Program (GULP) \citep{gale_gulp_1997}. The Buckingham-Coulomb pair potential \citep{buckingham_classical_1938} is commonly used for materials with an ionic character such as metal oxides. It has the form: 
\begin{equation}
    U(r_{ij})=\frac{q_iq_j}{r_{ij}}+A\ \mbox{exp}\left(-\frac{r_{ij}}{B}\right)-\frac{C}{r^6_{ij}},
    \label{eq:1}
\end{equation}
where $q_{i}$ and $q_j$ are the charges of atoms $i$ and $j$, $r_{ij}$ is the distance between them and \textit{A, B} and \textit{C} are the Buckingham pair parameters. The first term in Eq. \ref{eq:1} describes electrostatic interactions between the ions. The second term accounts for the Pauli principle and it represents the short range repulsion caused by the fact that ions occupy a certain volume in space. Lastly, the third term accounts for the attractive Van der Waals interaction. 
\begin{table*}[]
\begin{center}
\caption{\label{tab:buck_param}Parametrizations of the Buckingham-Coulomb pair potential used for the first stage in the optimization process of each cluster.}
\begin{tabular}{c cc c }
\cline{2-4}
                                                        & \multicolumn{2}{c}{{VO}}                         & {VO$_2$}   \\ \cline{2-4} 
                                                        & \multicolumn{1}{c}{Charge reduction} & Core-Shell model & Charge reduction\\ \hline
\multicolumn{1}{ c }{q (V) Core}                        & \multicolumn{1}{c}{1.6}              & 0.2              & 2.4                               \\
\multicolumn{1}{ c }{q (V) Shell}                       & \multicolumn{1}{c}{-}                & 1.8              & -                                 \\
\multicolumn{1}{ c }{q (O) Core}                        & \multicolumn{1}{c}{-1.6}             & 0.513            & -1.2                             \\
\multicolumn{1}{ c }{q (O) Shell}                       & \multicolumn{1}{c}{-}                & -2.513           & -                                 \\
\multicolumn{1}{ c }{A (V-O) {[}eV{]}}                  & \multicolumn{1}{c}{1946.254}         & 3000.000             & 3228.972                        \\
\multicolumn{1}{ c }{B (V-O) {[}$\AA${]}}               & \multicolumn{1}{c}{0.261}            & 0.3              & 0.261                           \\
\multicolumn{1}{ c }{C (V-O) {[}eV $\cdot \AA^{-6}${]}} & \multicolumn{1}{c}{0}                & 0                & 0                                \\
\multicolumn{1}{ c }{A(O-O) {[}eV{]}}                   & \multicolumn{1}{c}{25.41}            & 25.41            & 25.41                            \\
\multicolumn{1}{ c }{B(O-O) {[}$\AA${]}}                & \multicolumn{1}{c}{0.6937}           & 0.6937           & 0.6937                            \\
\multicolumn{1}{ c }{C(O-O) {[}eV $\cdot \AA^{-6}${]}}  & \multicolumn{1}{c}{32.32}            & 32.32            & 32.32                            \\
\multicolumn{1}{ c }{Spring (O)}                        & \multicolumn{1}{c}{-}                & 20.53            & -                               \\
\multicolumn{1}{ c }{Spring (V)}                        & \multicolumn{1}{c}{-}                & 300              & -                               \\ \hline
\end{tabular}
\end{center}
\end{table*}
Unlike for oxides of silicon, aluminum and titanium, there are sparse parameter sets for V-O systems available in the literature. Therefore, we used Buckingham pair parameters resembling those known for other metal oxides. We have applied different parametrizations of the Buckingham-Coulomb pair potential to optimize the seed structure geometries. A summary of the different parameters sets used in this study can be found in Table \ref{tab:buck_param}. As initial seed geometries we use a large number of polymorph structures reported for clusters of magnesium oxide \citep{chen_structures_2014}, aluminium oxide (\citep{gobrecht_global_2018}, and \citep{gobrecht_bottom-up_2022}), silicon oxide \citep{bromley_under_2016}, titanium oxide \citep{sindel_revisiting_2022}  and silicon carbide \citep{gobrecht_nucleation_2017}.\\
Since vanadium and oxygen have different electronegativities (3.44 and 1.63 on the Pauling scale \citep{pauling_nature_1932}) we have accounted for polarization effects in our calculations. For the vanadium monoxide isomers we applied two different parameter sets. An initial set of calculations was performed with reduced electrostatic charges for both the cations and the anions, and with \emph{A}, \emph{B}, and \emph{C} parameters for the V-O interactions from \cite{lewis_potential_1985} and for the O-O interactions from \cite{bush_self-consistent_1994}. A second set of calculations was performed using the core-shell model for each geometry. In the core–shell model \citep{dick_theory_1958}  the ion is divided into two closely lying coordinates connected by a spring with a spring constant $k$: a core that contains all the atomic mass, and a massless charged shell. For the second set of calculations the electrostatic charges were not reduced, but distributed between the core and the shell and shell-shell values were used for both the V-O (estimated based on the values available in \cite{lewis_potential_1985}) and O-O \citep{bush_self-consistent_1994} interactions. Both sets provided similar optimized geometries with slight differences in the bond lengths and the energetic ordering of the isomers. Since the core set produced more candidates that resulted on global minima after the DFT calculations (sizes N=2,4,5,6,9,10), and as previously mentioned, interatomic potential parameters for the core-shell model are not available for many systems, we decided to use the reduced charges method for vanadium oxide clusters with stoichiometries different from 1:1.

\subsection{Optimization with Density Functional Theory}
\label{sec:methods_DFT}

\begin{figure}
\centering
 \includegraphics[width=9.2cm]{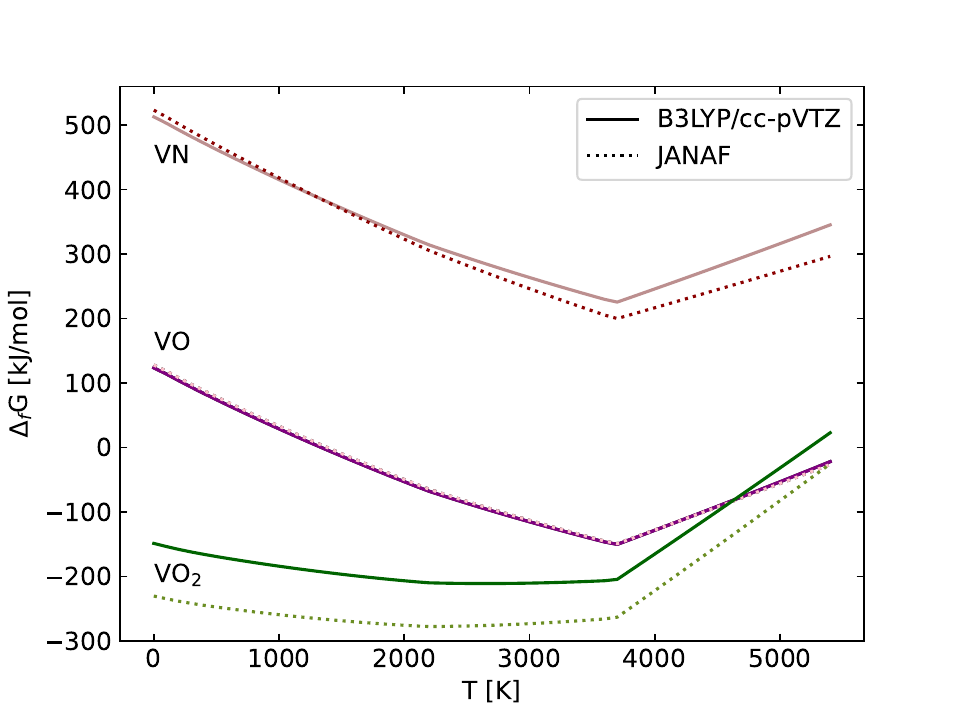}
 \caption{Gibbs free energies of formation (in kJ mol$^{-1}$) for VN, VO and VO$_{2}$ as a function of temperature. \textbf{Dotted lines} correspond to the values available on JANAF database and \textbf{full lines} represent the data obtained from our DFT calculations at the B3LYP/cc-pVTZ level of theory using the RRHO partition function.}
 \label{fig:comp_JANAF}
\end{figure}
The candidate structures generated with the methods detailed in Sect. \ref{sec:methods_FFM}  are subsequently optimized at the quantum level of theory with the computational chemistry software package \texttt{Gaussian16} \citep{frisch_gaussian16_2016}. We use the hybrid  B3LYP density functional \citep{becke_new_1993} as a compromise between accuracy and computational cost, which will be discussed later in this section. 
For our calculations we have used the correlation consistent polarized valence triple zeta basis set ( cc-pVTZ \cite{wilson_gaussian_1996}), whose basis functions are known to accurately describe electronic orbitals at an affordable computational cost. The accuracy of our choice was determined by comparing the computational results obtained from the DFT calculations with the experimental values published in the NIST-JANAF Thermochemical tables \footnote{https://janaf.nist.gov/} (Fig. \ref{fig:comp_JANAF}). \\
Generally, the binding energies of the vanadium oxides for a temperature of T=0 K are calculated as:
\begin{equation}
    E_{bind} (\mbox{V}_{x}\mbox{O}_{y})_{\text{N}}= E_{Pot} (\mbox{V}_{x}\mbox{O}_{y})_{\text{N}} - x \cdot {\text{N}} \cdot E_{Pot} (\mbox{V}) - y \cdot {\text{N}} \cdot E_{Pot} (\mbox{O})
    \label{eq:binding}
\end{equation}
For coherence and consistency, the potential energies are calculated at the B3LYP/cc-pVTZ level of theory showing very good agreement ($\sim$10 kJ mol$^{-1}$) with the experimental values for VO and a reasonable agreement ($\sim$22 kJ mol$^{-1}$) for VO$_2$. Moreover,  this functional/basis set was found to accurately predict the binding energies of other transition metal oxides like TiO$_2$ \citep{sindel_revisiting_2022}. For each global minima (GM) candidate we have also performed a frequency analysis with the same functional/basis set to determine if the candidate is a true minima and not a transition state and to calculate the relevant partition functions. The partition functions of the GM candidates are computed using the Rigid Rotor Harmonic Oscillator (RRHO) approximation, as implemented in the \texttt{thermo.pl} code \citep{irikura_thermopl_2002}. With the help of the partition functions the temperature-dependent thermodynamic potentials such as the entropy S, the enthalpy of formation, $\Delta_{f}$H(T), and the Gibbs free energy of formation $\Delta_{f}$G(T) can be derived (see Sect. \ref{sec:DFT_results} ).\\
The choice of the B3LYP/cc-pVTZ method is based on the replication of the NIST-JANAF values for the VO molecule, which can be seen on Fig. \ref{fig:comp_JANAF}.
The calibration was done for the vanadium monoxide (VO) molecule because we aim to get accurate energies for the first steps of cloud 
formation using a bottom-up approach. We know that in the bulk the preferred oxidation state is V$_{2}$O$_5$ (vanadium (V) oxide), corresponding to the thermodynamically most stable form of solid vanadium oxide \citep{yan_single-crystalline_2009}. Starting from atomic vanadium we study the intermediate oxidation steps, with the smallest molecule being diatomic VO. Therefore, accurate values for the energy, vibration frequency and rotational constant of VO are essential for modeling the vanadia nucleation starting from gas-phase molecules. In Fig.\ref{fig:comp_JANAF} we compare the Gibbs Free energies of formation ($\Delta_f$G(T)) for the vanadium-bearing gas-phase molecules for which data is available in the NIST-JANAF Thermochemical tables. The excellent agreement between the DFT results and the JANAF values for VO confirms that our choice of functional and basis set (B3LYP/cc-pVTZ) accurately represents the behavior of VO for the entire temperature range that we are studying. The agreement between our data and the JANAF table for VN is reasonable at lower temperatures and starts to diverge above $\sim$3000 K. The results for VO$_2$ follow the same trend but show an offset of about 80-100 kJ mol$^{-1}$ .
This discrepancy likely to arises from an erroneous enthalpy estimation based on studies of \cite{frantseva_mass-spectrometer_1969}. In fact, \cite{balducci_thermochemical_1983} revisited and reevaluated the VO$_2$ enthalpy in an experimental study and found significantly higher enthapies of formation. A comparison of the binding energies published in JANAF, the results of \cite{balducci_thermochemical_1983} and the findings of this study can be found in Table \ref{tb:binding_energies}. The VO bond energy derived from JANAF and B3LYP/cc-pVTZ calculations is within the confidence interval of the \cite{balducci_thermochemical_1983} values. This is not the case for VO$_2$, but the results obtained with our calibrated DFT calculations are considerably closer to the results of \cite{balducci_thermochemical_1983} than those reported in the JANAF table.

\begin{table}[]
\caption{VO and VO$_2$ electronic binding energies from different sources. Uncertainties for the JANAF values are assumed to be the chemical accuracy of 4 kJ mol$^{-1}$ \citep{chase_nist-janaf_1998} whereas uncertainties for the \cite{balducci_thermochemical_1983} values are 18 kJ mol$^{-1}$.}
\begin{center}
\begin{tabular}{c c c }
\cline{2-3}
                                            & E$_{bind}$ VO  & E$_{bind}$ VO$_2$  \\
                                            & {[}kJ mol$^{-1}${]}               & {[}kJ mol$^{-1}${]}                \\ \hline
\multicolumn{1}{ c }{B3LYP/cc-pVTZ}                   & 635.7                      & 1154.7                      \\
\multicolumn{1}{ c }{JANAF}                 & 631.2                   & 1236.3                     \\
\multicolumn{1}{ c }{\citep{balducci_thermochemical_1983}} & 625.5                      & 1177.0                        \\ \hline
\end{tabular}
\label{tb:binding_energies}
\end{center}
\end{table}
\noindent Owing to the open shell character of the vanadium atom with an electronic ground state  of $^{4}F_{3/2}$ and the VO molecule with ground state of $^{4}\Sigma^{-}$, also other V-bearing molecules and clusters can have spin quantum numbers larger than one, i.e. singlets. Therefore, we investigated several spin multiplicities for each of our vanadium cluster candidates.

\subsection{Implementation of data for equilibrium chemistry calculations}
\label{sec:methods_Eq.Chemistry}

In the following we will use the term abundance defined as species concentration relative to the total gas density. The thermodynamic ranges for which each vanadium bearing species is the most abundant in a given substellar atmosphere can be determined with chemical equilibrium calculations. The abundances obtained from chemical equilibrium can later be used as a starting point to calculate reaction rates in a kinetic approach.\\
We applied the gas-phase equilibrium code \texttt{GGChem} from \cite{woitke_equilibrium_2017} to obtain the equilibrium abundances. 
\texttt{GGChem} applies the law of mass action derived by minimizing the Gibbs free energy of the total gas ($\Delta_f$G$^\plimsoll$) to calculate the number densities of each species at a given temperature and pressure. In order to implement the DFT data into the code we first calculated the Gibbs free energies of dissociation of each species ($\Delta_rG_i^\plimsoll$(T)) following the approach from \cite{woitke_equilibrium_2017} and \cite{stock_fastchem_2018}:
\begin{equation}
    \Delta_rG_i^\plimsoll \textrm{(T)} = \Delta_fG_i^\plimsoll \textrm{(T)} - \sum_{j \in \epsilon_0}v_{ij}  \Delta_fG_j^\plimsoll \textrm{(T)}
    \label{eq:gibbs_dissociation}
\end{equation}
where $\Delta_fG_i^\plimsoll$(T) is the Gibbs free energy of formation of species $i$. The second term on the right hand side represents the sum over the Gibbs free energies of formation of each element in the species. $\epsilon_0$  is the set of all the elements included in the code and $v_{ij}$ the stoichiometric coefficients of the elements within species $i$. Since the natural logarithm of the dimensionless equilibrium constant ($\bar{K_i}$(T)) of a certain species can be defined as:
\begin{equation}
    \textrm{ln}\ \bar{K_i} \textrm{(T)}= - \frac{\Delta_rG_i^\plimsoll\textrm{(T)}}{R\textrm{T}}
\end{equation}
We can express $\Delta_rG_i^\plimsoll$(T) as:
\begin{equation}
    \Delta_rG_i^\plimsoll \textrm{(T)} = - R\textrm{T}\ \textrm{ln}\ (\bar{K_i})
\end{equation}
and fit the Gibbs free energies of dissociation from Eq. \ref{eq:gibbs_dissociation} with an expression in temperature with 5 coefficients ($a_0,a_1,b_0,b_1,b_2$) used as input into \texttt{GGChem}:
\begin{equation}
     \Delta_rG_i^\plimsoll \textrm{(T)} = - R\textrm{T}\ \left( \frac{a_0}{\textrm{T}} + a_1\textrm{ln}\textrm{(T)}+ b_0+b_1\textrm{T}+b_2\textrm{T}^2 \right)
     \label{eq:fit}
\end{equation}
All the calculations are done under the assumption of local thermal equilibrium, therefore the gas temperature and the cluster temperatures are the same.

\section{Results}
\label{sect:results}

There are several thermochemical properties of interest for astrophysical studies, including the Gibbs free energies of formation, the change in enthalpy or the entropy. 
The following sections present the geometries and derived thermochemical properties of all the vanadium oxide clusters covered in this study. The thermochemical properties are then used to evaluate the impact of adding clusters on the gas-phase abundances of all vanadium bearing species. The gas-phase abundances in chemical and thermal equilibrium can be used as a starting point in the modelling of atmosphere kinetics, which is needed in order to interpret observational data (e.g., James Webb Space Telescope (JWST)).

\subsection{Lowest energy cluster structures}
\begin{figure*}[hbt!]
\centering
\resizebox{19cm}{!}{
\begin{tabular}{ccccc}
\includegraphics[width=0.19\textwidth]{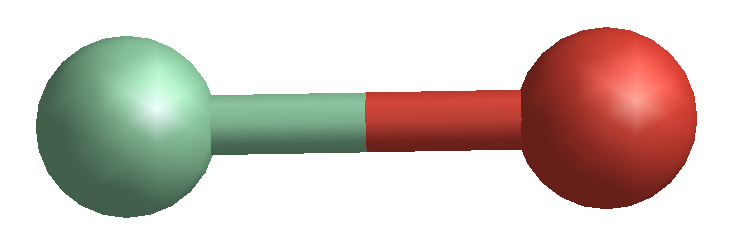}          & \includegraphics[width=0.19\textwidth]{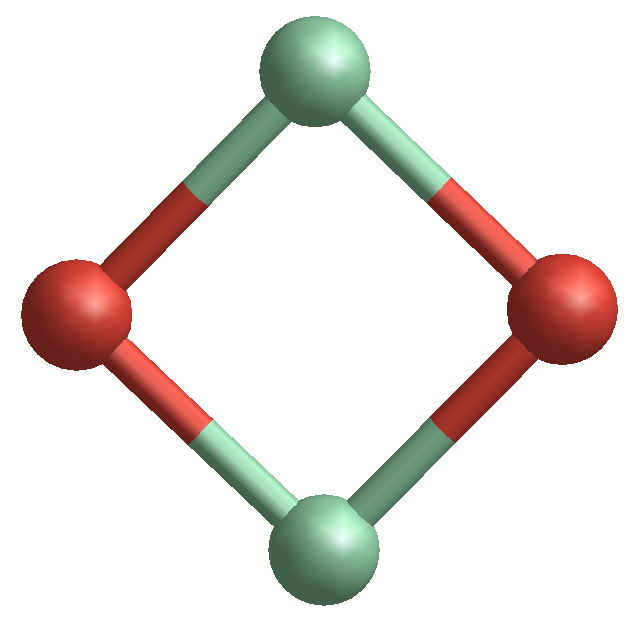}           & \includegraphics[width=0.19\textwidth]{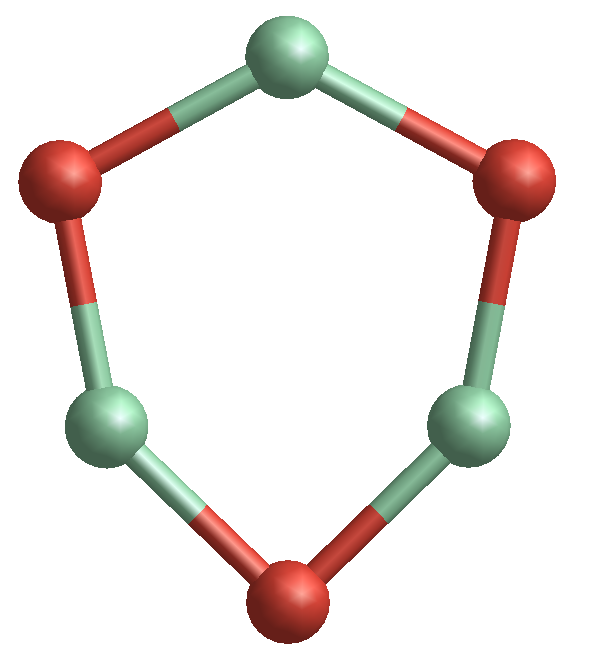}           & \includegraphics[width=0.19\textwidth,angle=90]{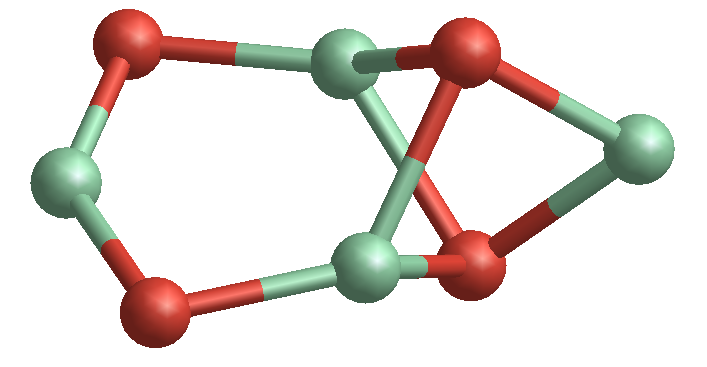}    \vspace{0.5cm}    & \includegraphics[width=0.19\textwidth,angle=90]{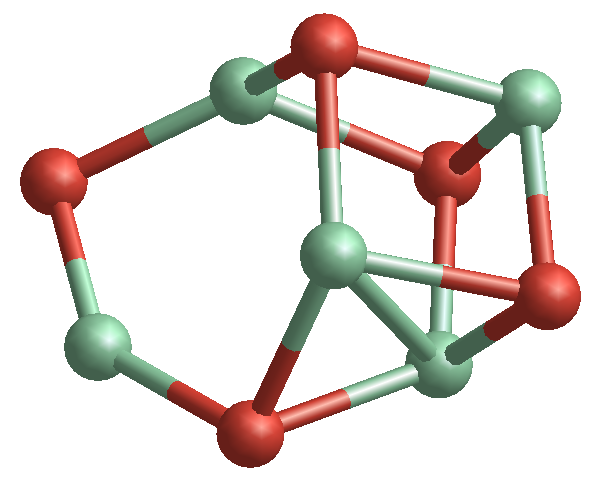}             \\
(a) VO         & (b) V$_2$O$_2$    & (c) V$_3$O$_3$   & (d) V$_4$O$_4$  & (e) V$_5$O$_5$      \\
\includegraphics[width=0.19\textwidth]{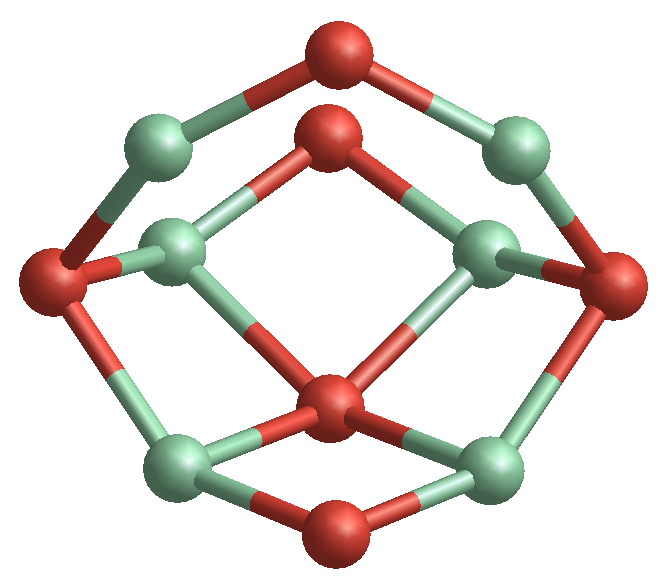}              & \includegraphics[width=0.19\textwidth,angle=90]{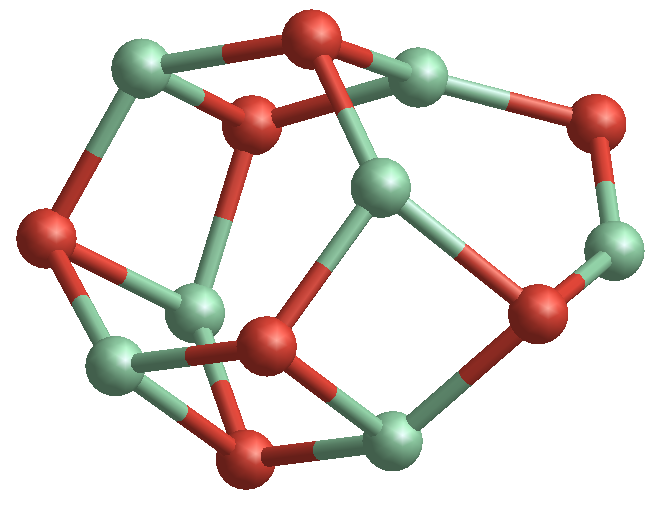}            & \includegraphics[width=0.19\textwidth]{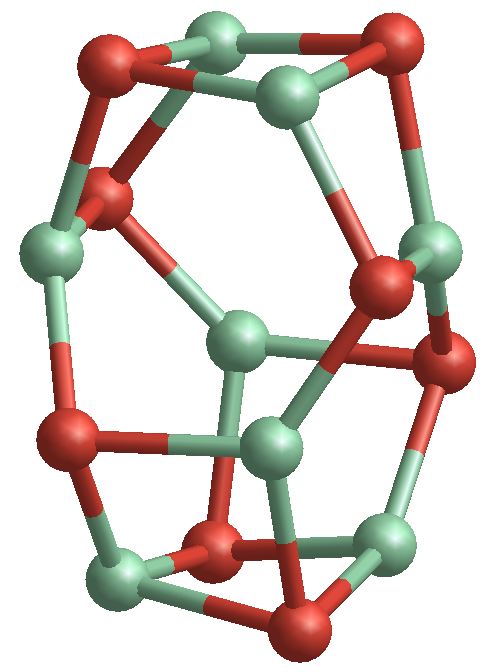}            & \includegraphics[width=0.19\textwidth]{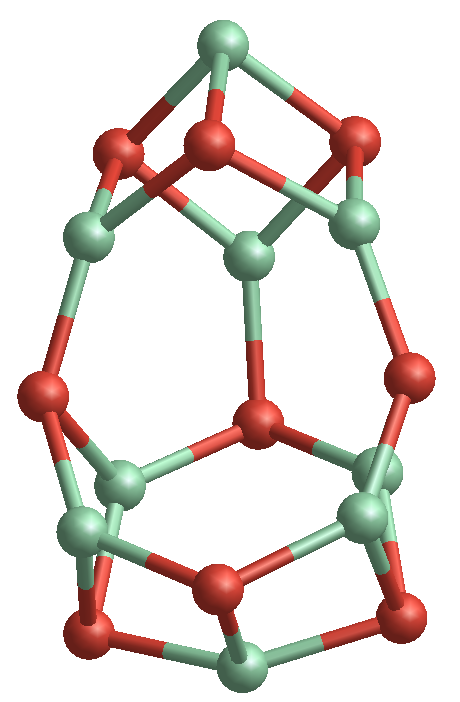}       \vspace{0.2cm}     & \includegraphics[width=0.19\textwidth,angle=90]{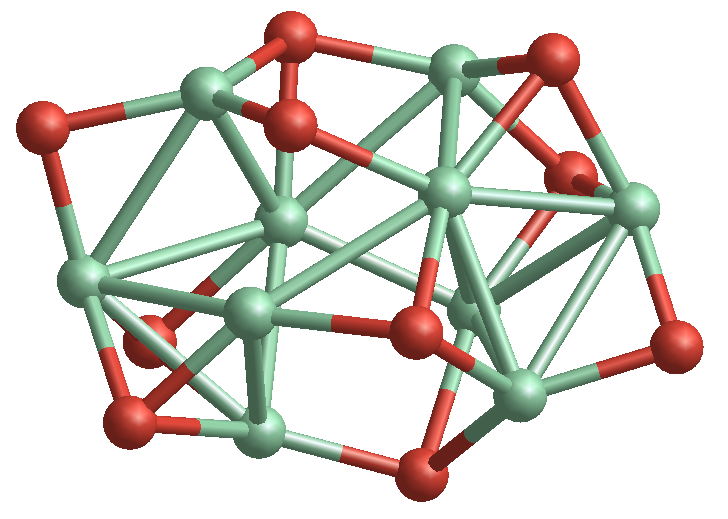}              \\
(f) V$_6$O$_6$ & (g) V$_7$O$_7$ & (h) V$_8$O$_8$ & (i) V$_9$O$_9$ & (j) V$_{10}$O$_{10}$ \\
\includegraphics[width=0.19\textwidth]{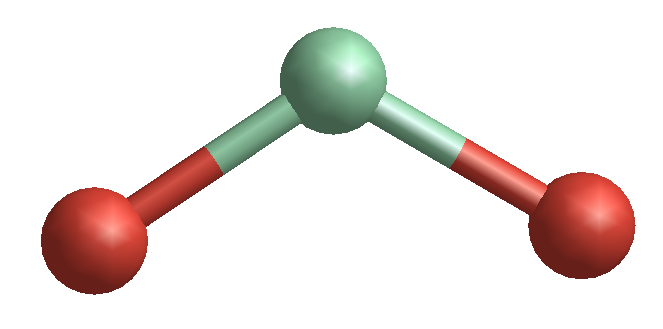}          & \includegraphics[width=0.19\textwidth]{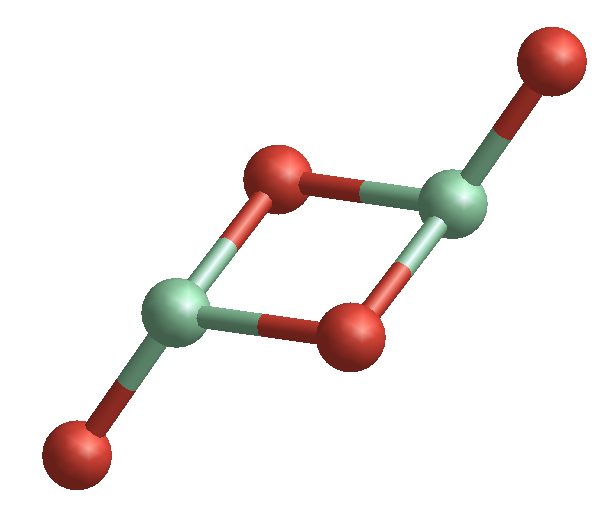}           & \includegraphics[width=0.19\textwidth]{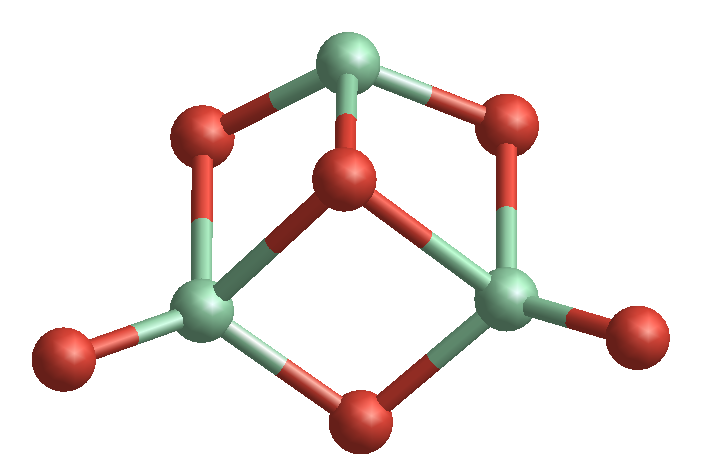}           & \includegraphics[width=0.19\textwidth]{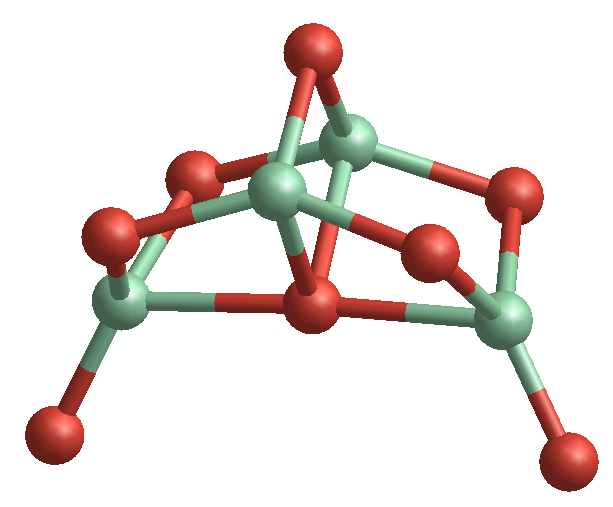}    & \includegraphics[width=0.19\textwidth]{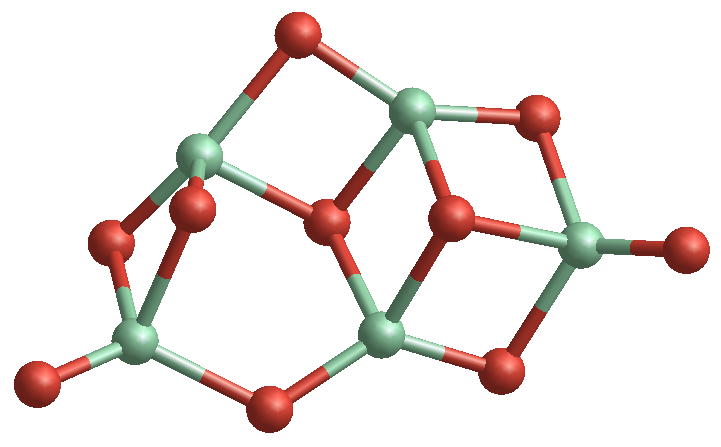}             \\
(k) VO$_2$          & (l) V$_2$O$_4$     & (m) V$_3$O$_6$   & (n) V$_4$O$_8$    & (o) V$_5$O$_{10}$     \\
\includegraphics[width=0.19\textwidth]{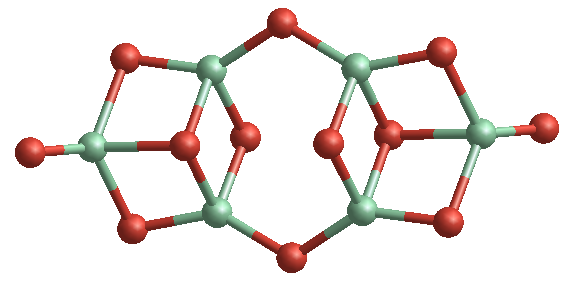}              & \includegraphics[width=0.19\textwidth]{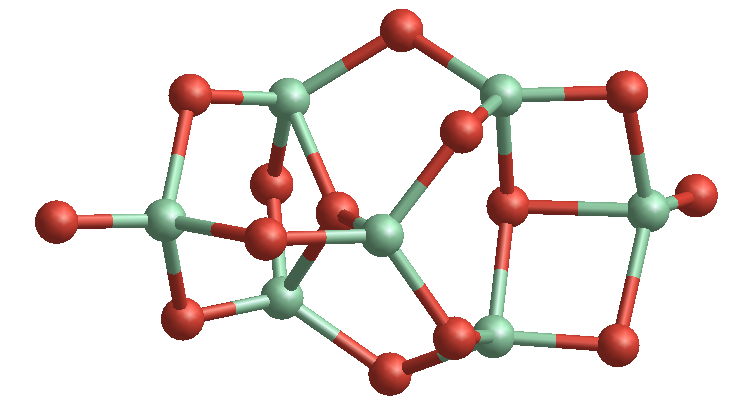}            & \includegraphics[width=0.19\textwidth]{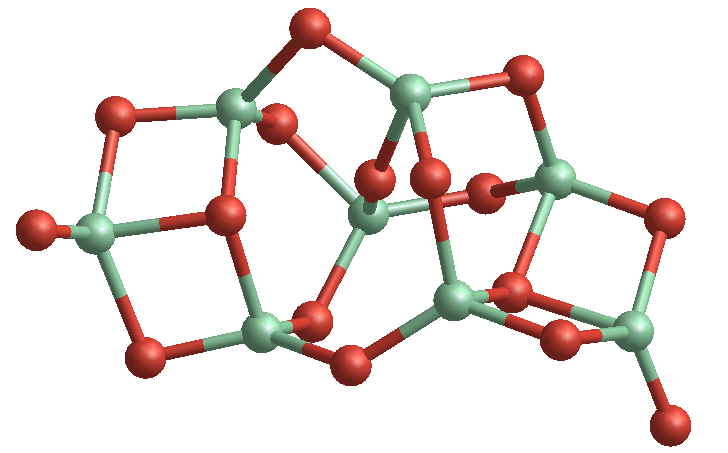}            & \includegraphics[width=0.19\textwidth]{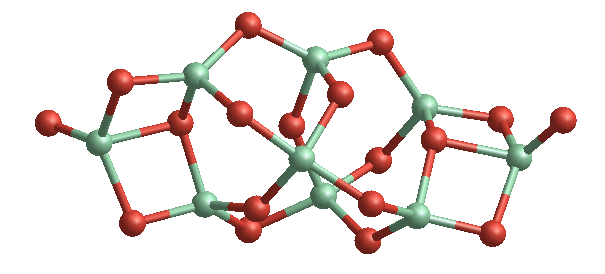}            & \includegraphics[width=0.19\textwidth]{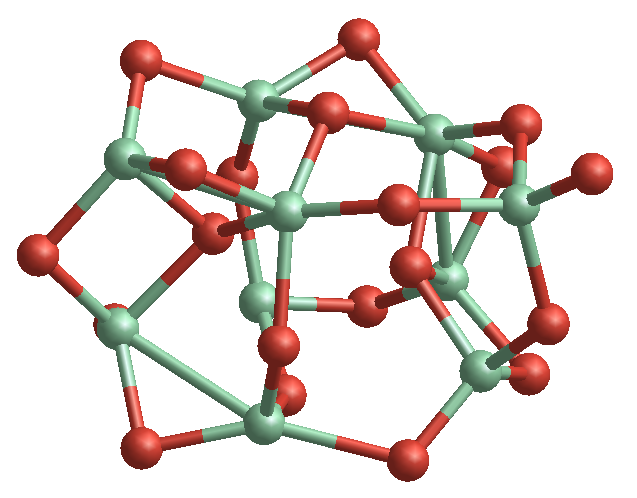}              \\
(p) V$_6$O$_{12}$ & (q) V$_7$O$_{14}$ & (r) V$_8$O$_{16}$ & (s) V$_9$O$_{18}$ & (t) V$_{10}$O$_{20}$ \\
\includegraphics[width=0.19\textwidth]{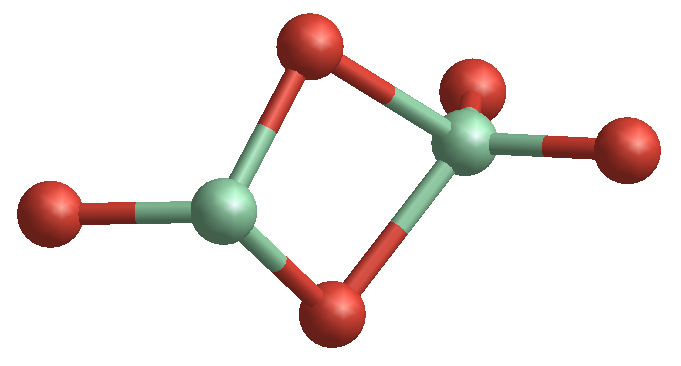}          & \includegraphics[width=0.19\textwidth]{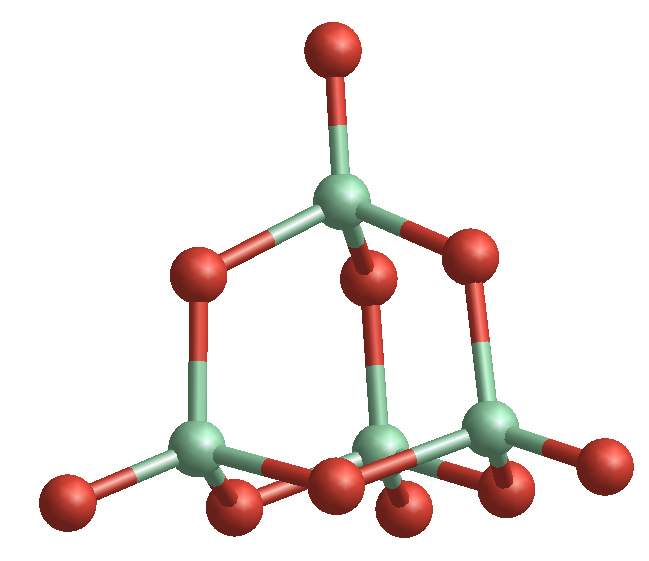}           & \includegraphics[width=0.19\textwidth]{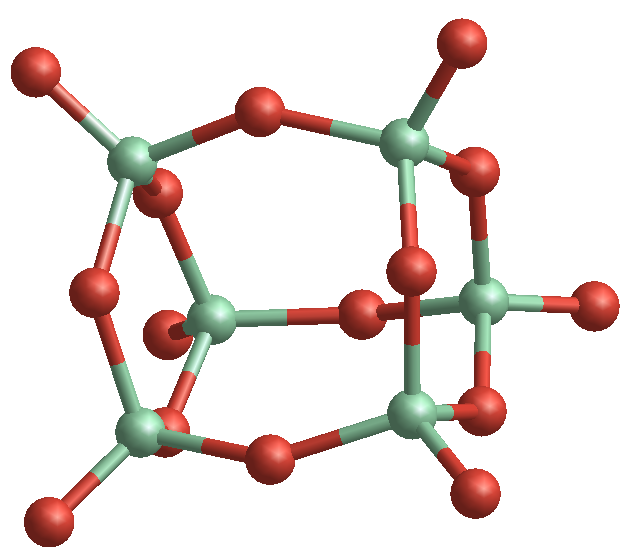}           & \includegraphics[width=0.19\textwidth]{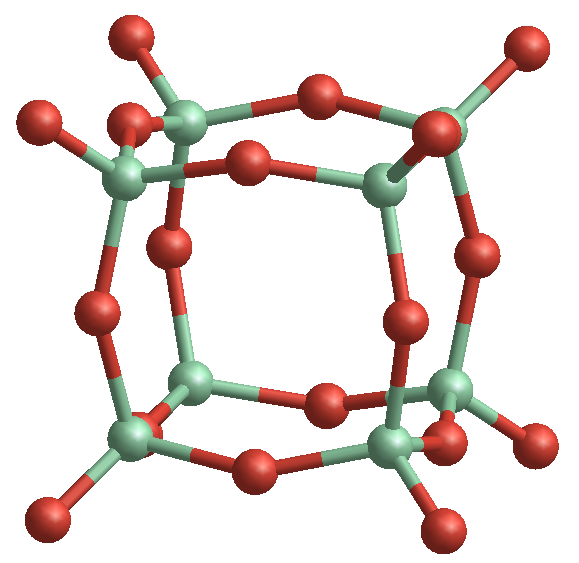}              \\
(u) V$_2$O$_5$       \vspace{0.1cm}      & (v) V$_4$O$_{10}$     \vspace{0.1cm}& (w) V$_6$O$_{15}$    \vspace{0.1cm} & (x) V$_8$O$_{20}$     \vspace{0.1cm} &
\end{tabular}}
\caption{Geometries of the global minima candidates for (VO)\textsubscript{N} and (VO$_2$)\textsubscript{N} with N = 1-10 and (V$_2$O$_5$)\textsubscript{N} with N = 1-4. Vanadium and oxygen atoms are shown in green and red, respectively.}
\label{fig:geometries}
\end{figure*}

For each considered vanadium oxide compound the 10 -100 lowest-energy candidate isomers found with the interatomic pair potential (see Sect. \ref{sec:methods_FFM}) were subsequently optimized at the B3LYP/cc-pVTZ level of theory with \texttt{Gaussian16} \citep{frisch_gaussian16_2016}. 
The geometries of the Global Minima (GM) candidates for each considered stoichiometry and cluster size are presented in Fig. \ref{fig:geometries}.

\noindent In the following the properties of the lowest-energy V\textsubscript{x}O\textsubscript{y} isomers listed in Table \ref{GM_table} are described in detail. We report the electronic binding energy of the GM clusters at T=0 K as calculated with the B3LYP/cc-pVTZ method, the spin multiplicity M$_s$, the point symmetry group in the Sch\"onflies notation, and the average V-O bond length in \AA.
Vanadium has atomic number 23 and therefore at least one unpaired electron. Its electronic ground state is a quartet. As a consequence clusters containing an odd number of V cations show even spin multiplicities with values M$_S$=2,4,6,..., and vice versa, clusters with an even number of V atoms show an odd spin multiplicity.\\ 
The (VO)\textsubscript{N} cluster show high spin multiplicities, which is a consequence of different valencies of vanadium and oxygen.  
The lowest-energy state of the VO molecule (Fig. \ref{fig:geometries}(a)) is a quartet state, i.e. spin multiplicity of 4. 
We find a bond dissociation energy of 636 kJ mol$^{-1}$ which is close to the experimental value of 621 kJ mol$^{-1}$ derived by \citet{jakubikova_density_2007}. They find a bond length of 1.61 \AA{}, which is about 0.03 \AA{} longer than the optimized distance in our study. 
The lowest-energy V$_2$O$_2$ cluster structure (Fig. \ref{fig:geometries}(b)) was extensively investigated by  
\citet{hubner_multiple_2017}. The authors use Multi Reference Configuration Interaction (MRCI) 
calculations and report a singlet electronic ground state and multiple vanadium–vanadium bonds, which 
is in contrast to previous findings \citep{jakubikova_density_2007}. 
With our single reference B3LYP/cc-pVTZ calculations we find a septet (M$_s$=7) state and no V-V bond 
as the lowest-energy configuration, consistent with the results of \citet{jakubikova_density_2007}.
For the corresponding triplet state, we find an energy that is only 25 kJ mol$^{-1}$ above the reported M$_s$=7 state.
High level test calculations indicate indeed a multiconfigurational character of the V$_2$O$_2$ global 
minimum. However, owing to the high computational cost an investigation with such post-Hartree-Fock 
methods is limited to relatively small systems and prohibitive for the larger cluster sizes considered in 
this study.   
The lowest-energy V$_3$O$_3$ isomer (Fig.\ref{fig:geometries}(c)) is a closed ring structure with non-uniform bond angles as reported in \citet{kaur_understanding_2019}, which is similar to the GM candidates of Si$_3$O$_3$ studied by \citep{bromley_under_2016}. 
The structures of larger (VO)\textsubscript{N} with N$\ge$4 were not investigated in previous studies 
and their respective GM candidates are reported for the first time in our present study.
For sizes N=4-9 (Fig. \ref{fig:geometries}(d-i)) we find void cages with alternating V-O ordering as the most favourable isomers. 
For N=4,5,6 (Fig.\ref{fig:geometries}(d-f)) and 9 (Fig. \ref{fig:geometries}(i)) the GM candidates are symmetric structures showing a mirror plane.
(VO)$_{10}$ (Fig. \ref{fig:geometries}(j)) represents an outlier in this series of GM candidates as it exhibits several V-V bonds, is 
a singlet state, and is very compact, in contrast to smaller sized (VO)$_{9}$.\\
Some of the (VO$_2$)\textsubscript{N} GM candidates have geometries equivalent to the GM isomers of (TiO$_2$)\textsubscript{N} reported in 
\cite{lamiel-garcia_predicting_2017} and \cite{sindel_revisiting_2022}. They include N=1,2,3,4, and 7 (Fig. \ref{fig:geometries}(k-n,p)). 
Compared to the TiO$_2$ monomer the VO$_2$ (Fig. \ref{fig:geometries}(k)) molecule exhibits a similar geometry with slightly shorter 
bond lengths (1.61 \AA{}) and a slightly larger bond angle (116$^\circ$).
The VO$_2$ dimer (Fig. \ref{fig:geometries}(l)), i.e. V$_2$O$_4$, shows a \textit{trans} configuration. In 
contrast to \citet{jakubikova_density_2007} we do not find a broken symmetry of this singlet GM candidate.
To our knowledge the GM candidate for V$_3$O$_6$ (Fig. \ref{fig:geometries}(m)) is reported for the first time. 
The planar V$_3$O$_6$ structure reported in \citet{jakubikova_density_2007} and \cite{kaur_understanding_2019} lies 178 kJ mol$^{-1}$ above 
our GM candidate and corresponds to a transition state with an imaginary frequency. 
Also the GM candidate of V$_4$O$_8$ (Fig. \ref{fig:geometries}(n)) shows a similar geometry to the corresponding Ti$_4$O$_8$ GM 
candidate.
For the isomer reported by \citet{kaur_understanding_2019} we find an energy 55 kJ mol$^{-1}$ above our GM candidate and two imaginary frequencies.  
The structures of the (VO$_2$)\textsubscript{N}, N$=$4-10 (Fig. \ref{fig:geometries}(n-t)), GM candidates represent hitherto unreported structures. 
Apart from  (VO$_2$)$_{10}$ (Fig. \ref{fig:geometries}(t)) they are characterised by two singly coordinated terminal oxygen atoms with 
comparatively short bond lengths. Moreover, these favourable cluster isomers are not symmetric and 
belong to the C$_{1}$ point group with the exception of  (VO$_2$)$_{6}$ (Fig. \ref{fig:geometries}(p)).\\
The GM candidates of the (V$_2$O$_5$)\textsubscript{N}, N=1-4 (Fig. \ref{fig:geometries}(u-x)), cluster family show symmetric structures and are singlet states. As members of the vanadia stoichiometry like solid V$_2$O$_5$, these GM candidate 
clusters are well studied and were previously reported in e.g. \citet{vyboishchikov_gas-phase_2000, lasserus_vanadiumv_2019}. Our search for low-energy (V$_2$O$_5$)\textsubscript{N} isomers confirms the results of these studies. The atomic coordinates of the reported GM cluster candidates are available in electronic form at the CDS.

\begin{table}[hbt!]
\caption{Properties of the (VO)\textsubscript{N}, (VO$_2$)\textsubscript{N}, N=1-10, and (V$_2$O$_5$)\textsubscript{N}, N=1-4, global minima candidates listing (1) cluster size, (2)  the electronic binding energy E$_{b}$ (in kJ mol$^{-1}$), (3) the electric binding energy normalized to the cluster size E$_{b}$/N (in kJ mol$^{-1}$), (4) the spin multiplicity M$_s$, (5) point group symmetry in the Sch\"onflies notation, (6) Average VO bond length  $\overline{d}$(VO) in \AA \label{GM_table}.}
\begin{tabular}{l r c c c c }
N & E$_{b}$ (kJ mol$^{-1}$) & E$_{b}$/N  & M$_s$            & Sym & $\overline{d}$(VO) (\AA{})   \\
\hline
(VO)\textsubscript{N} \\
\hline
1      & 636        &   636.0&  4       & C$_{\infty v}$       & 1.579  \\
2      & 1497       &   748.5&  7       & C$_{2v}$       & 1.840  \\
3      & 2417       &   805.6&  4       & C$_{2v}$       & 1.814  \\
4      & 3320       &   830.0&  5       & C$_{2v}$       & 1.901  \\
5      & 4280       &   856.0&  4       & C$_{s}$        & 1.943       \\
6      & 5314       &   885.6&  7       & C$_{s}$          & 1.940   \\
7      & 6393       &   913.3&  4       & C$_{1}$        & 1.945      \\
8      & 7253       &   906.6&  3       & C$_{1}$        & 1.951  \\
9      & 8180       &   908.8&  4       & C$_{s}$        & 1.959  \\
10     & 9014       &   901.4&  1       & C$_{1}$        & 1.931 \\
\hline
(VO$_2$)\textsubscript{N} \\
\hline
1      & 1150       &   1150.0& 2        & C$_{2v}$       & 1.612 \\
2      & 2671       &   1335.5& 1        & C$_{2v}$       & 1.729 \\
3      & 4330       &   1443.3&  2       & C$_{s}$        & 1.828  \\
4      & 5901       &   1475.3&  5       & C$_{2v}$       & 1.841  \\
5      & 7460       &   1492.0&  4       & C$_{1}$        & 1.841  \\
6      & 9055       &   1509.2&  5       & C$_{s}$        & 1.829  \\
7      & 10670      &   1524.3&  6       & C$_{s}$        & 1.826  \\
8      & 12177      &   1522.1&  5       & C$_{1}$        & 1.826  \\
9      & 13706      &   1522.8&  8       & C$_{1}$        & 1.819  \\
10     & 14907      &   1490.7&  1       & C$_{1}$        & 1.832  \\
\hline
(V$_2$O$_5$)\textsubscript{N} \\
\hline
1      & 3198      &    3198.0&  1        & C$_{s}$       & 1.731  \\
2      & 7025      &    3512.5&  1       & T$_{d}$       & 1.740  \\
3      & 10670     &    3556.6&  1       & C$_{2v}$     & 1.731  \\
4      & 14315     &    3578.8&  1       & T$_{d}$      & 1.725  \\
\end{tabular}
\label{tb:energies}
\end{table}

\subsection{Thermochemical properties}
\label{sec:DFT_results}

\begin{figure}[h!]
\centering
\includegraphics[width=9.5cm]{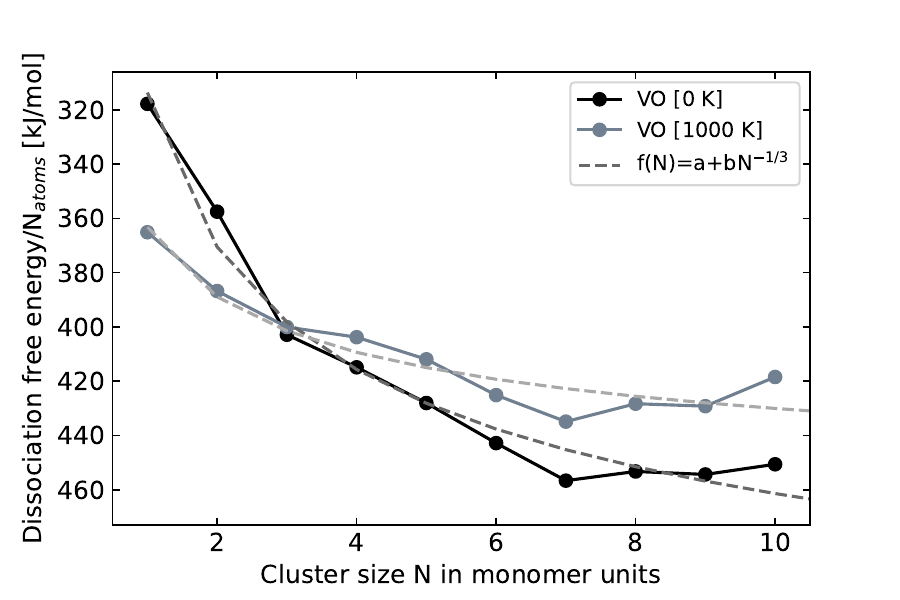}
\includegraphics[width=9.5cm]{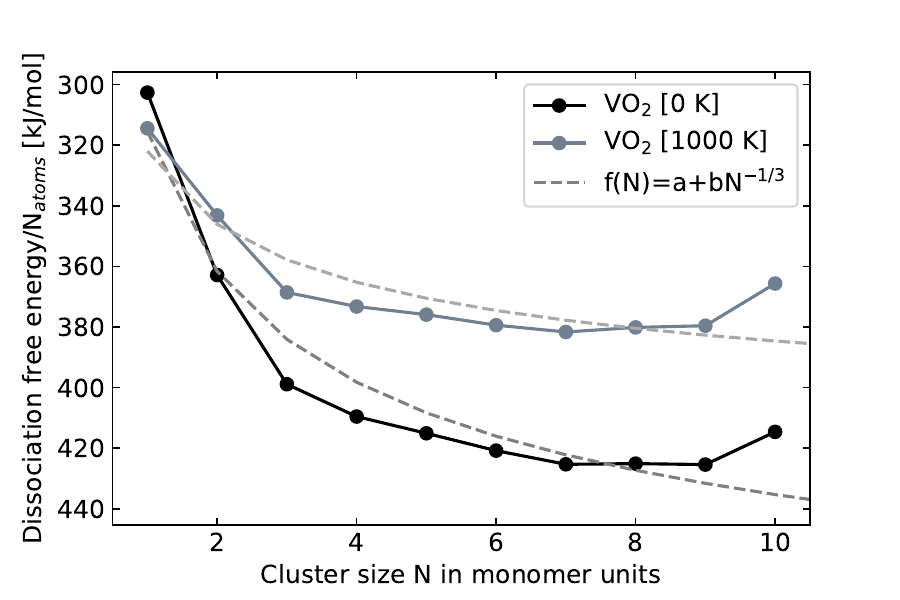}
\includegraphics[width=9.5cm]{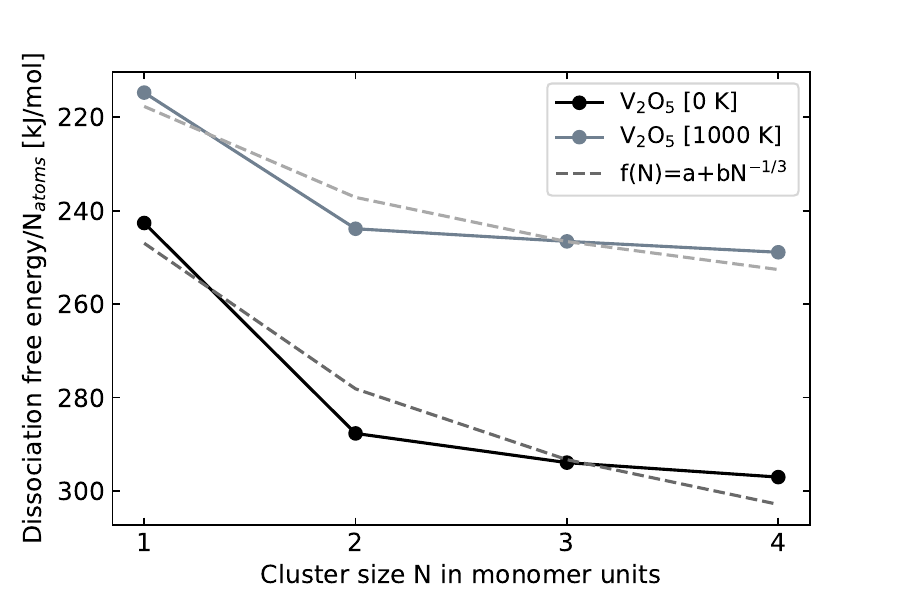}
  \caption{Dissociation free energy of each global (VO)\textsubscript{N} (\textbf{top}), (VO$_2$)\textsubscript{N} (\textbf{middle}) and (V$_2$O$_5$)\textsubscript{N} (\textbf{bottom}) minima candidate normalized by the cluster size at T=0 K and T=1000 K as a function of cluster size N and fitted to the Spherical Cluster Approximation as described in Eq. \ref{eq:spherical_approx}. }
\label{fig:binding_energy}
\end{figure}

\begin{table*}[]
\begin{center}
\caption{Summary of the thermochemical quantities (temperature T, entropy S, heat capacity C$_P$, enthalpy change
H(T)-H(0), enthalpy of formation dH$_f$ and Gibbs free energy of formation dG$_f$) obtained for each of our global minima candidates at T=1000 K \label{tb:thermo_data}.}
\begin{tabular}{ccccccc}
\hline
N                & S             & Cp            & H(T)-H(0)    & dHf          & dGf          & log Kf  \\
                 & {[}J/mol.K{]} & {[}J/mol.K{]} & {[}kJ/mol{]} & {[}kJ/mol{]} & {[}kJ/mol{]} &         \\ \hline
(VO)\textsubscript{N}         &               &               &              &              &              &         \\ \hline
1                & 271.245       & 36.011        & 32.675       & 116.174      & 28.679       & -1.498  \\
2                & 405.399       & 80.624        & 67.110       & 8.800        & -29.099      & 1.520   \\
3                & 522.554       & 129.349       & 107.760      & -152.132     & -123.436     & 6.448   \\
4                & 634.714       & 178.533       & 147.725      & -295.160     & -194.874     & 10.179  \\
5                & 747.347       & 227.625       & 188.721      & -495.488     & -324.085     & 16.928  \\
6                & 881.093       & 276.247       & 229.728      & -769.518     & -548.111     & 28.630  \\
7                & 971.451       & 325.040       & 268.293      & -1090.860    & -776.061     & 40.537  \\
8                & 1059.705      & 374.139       & 307.700      & -1192.140    & -781.845     & 40.839  \\
9                & 1192.355      & 423.278       & 349.791      & -1357.400    & -896.005     & 46.802  \\
10               & 1170.576      & 469.313       & 374.633      & -1447.606    & -780.682     & 40.778  \\ \hline
(VO$_2$)\textsubscript{N}     &               &               &              &              &              &         \\ \hline
1                & 332.902       & 55.452        & 47.494       & -157.093     & -184.456     & 9.635   \\
2                & 485.018       & 127.254       & 102.978      & -667.074     & -541.014     & 28.259  \\
3                & 635.445       & 199.212       & 158.021      & -1321.398    & -1040.226    & 54.335  \\
4                & 777.473       & 271.425       & 213.027      & -1887.303    & -1442.620    & 75.354  \\
5                & 932.913       & 343.433       & 270.133      & -2438.288    & -1843.506    & 96.294  \\
6                & 1080.234      & 415.435       & 325.255      & -3028.434    & -2275.434    & 118.855 \\
7                & 1216.479      & 488.064       & 383.216      & -3623.433    & -2701.139    & 141.091 \\
8                & 1358.515      & 560.124       & 439.064      & -4138.499    & -3052.702    & 159.455 \\
9                & 1509.882      & 632.901       & 496.576      & -4657.649    & -3417.680    & 178.487        \\
10               & 1580.892      & 705.182       & 548.898      & -4854.261    & -3379.763    &     176.584    \\ \hline
(V$_2$O$_5$)\textsubscript{N} &               &               &              &              &              &         \\ \hline
1                & 529.838       & 150.046       & 119.009      & -947.301     & -744.272     & 38.876  \\
2                & 846.462       & 317.406       & 247.232      & -2514.842    & -1895.570    & 99.013  \\
3                & 1201.915      & 483.753       & 377.540      & -3897.615    & -2900.929    & 151.527 \\
4                & 1588.065      & 650.380       & 510.351      & -5277.885    & -3934.482    & 205.514 \\ \hline
\end{tabular}
\end{center}
\end{table*}

\begin{table}[]
\caption{Summary of the fit parameters from Eq. \ref{eq:fit} for VO, VO$_2$ and V$_2$O$_5$ at T= 0 K and T= 1000 K.}
\begin{tabular}{ccccc}
\cline{2-5}
           & \multicolumn{2}{c}{T = 0 K}     & \multicolumn{2}{c}{T = 1000 K}  \\  
           & a {[}kJ/mol{]} & b {[}kJ/mol{]} & a {[}kJ/mol{]} & b {[}kJ/mol{]} \\ \hline
VO         & 589.4        & -275.7       & 487.9        & -124.7       \\
VO$_2$     & 539.1        & -223.5     & 438.9        & -116.9       \\
V$_2$O$_5$ & 397.9        & -151.0       & 312.0       & -94.3        \\ \hline
\end{tabular}
\label{tb:fit_parameters}
\end{table}

The spherical cluster approximation \citep{johnston_atomic_2002} was used to fit the potential energies of the global minima candidates as a function of the cluster size N allowing for an extrapolation to cluster sizes that are prohibitive to study with DFT.
Using the spherical cluster approximation it can be shown that the surface area to "bulk" ratio of the cluster is proportional to N$^{-1/3}$. This implies that many properties of the cluster, such as its dissociation energy or melting temperatures, can be approximately fitted by the following 
scaling law:
\begin{equation}
    f(N)= a\ + b\text{N}^{-1/3}
    \label{eq:spherical_approx}
\end{equation}
Where $a$ corresponds to the value of the property being studied but at the bulk phase of the material \citep{vines_size_2017}. 
The results of fitting the
global minima candidates for (VO)\textsubscript{N}, (VO$_2$)\textsubscript{N} with N=1-10 and (V$_2$O$_5$)\textsubscript{N} with N=1-4 are 
shown in Fig. \ref{fig:binding_energy} and a summary of the fit parameters can be found in Table 
\ref{tb:fit_parameters}. For (VO)\textsubscript{N} and (VO$_2$)\textsubscript{N} we obtain an acceptable agreement between the 
spherical cluster fit and our data for the smaller sizes but for the larger clusters the normalized energies cease to decrease monotonically and even can follow an opposite trend. 
This is the case for cluster sizes N=7-10 for (VO)\textsubscript{N} and (VO$_2$)\textsubscript{N} indicating that our candidate might not correspond to the lowest energy configuration. However, with our extensive searches of a great structural diversity we hope to have minimized the probability to miss a particularly favourable isomer.
The deviation from the fit may be due to the 1:1 and 1:2 stoichiometric ratios of V:O not being preferred as the clusters increase in size since the preferred oxidation state in the bulk is V$_2$O$_5$.\\
From the partition functions that include vibrational contributions we calculated the relevant thermochemical quantities as described in Sect. \ref{sec:methods_DFT}. We have summarized our results for T=1000 K in Table \ref{tb:thermo_data} and the complete tables (for the range 0 K$\leq$ T $\leq$ 5500 K) for each global minima candidate will be available electronically as supplementary material at the CDS. 

\subsection{Impact on abundances of vanadium-bearing species}

\begin{figure}
\centering
\includegraphics[width=9.4cm]{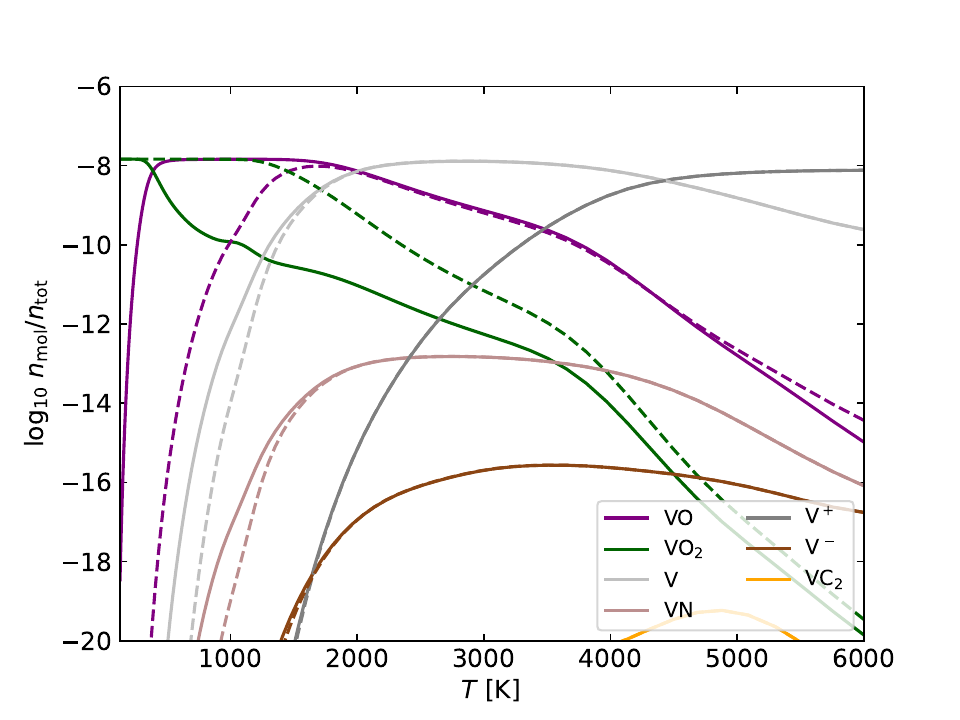}
  \caption{Comparison of the abundances of the vanadium bearing species currently present in \texttt{GGchem} as a function of temperature. The \textbf{dashed lines} correspond to the data from previous studies already implement in \texttt{GGchem} and the \textbf{full lines} correspond to the abundances obtained after implementing the $\Delta_f$G values obtained for VO and VO$_2$ from our DFT calculations. The calculations were run at $P=$ 1 bar and with solar elemental abundances from \cite{asplund_chemical_2009}.}
\label{fig:comparison_GGchem}
\end{figure}

As a first step we determine the impact of implementing the revised and more accurate thermochemical data derived in this study for the vanadium-bearing species already present in \texttt{GGChem}. 
Our results for VO$_2$ differ from the JANAF free energies \footnote{https://janaf.nist.gov/tables/O-076.html} presently used in \texttt{GGchem} as shown in Fig. \ref{fig:comp_JANAF} and hence differences in the equilibrium abundances are  expected. \\
As can be seen in Fig. \ref{fig:comparison_GGchem}, substituting the current $\Delta_f$G values for VO and VO$_2$ with the updated energies obtained in this study has a strong impact at lower temperatures (T $\lesssim$ 2000 K) for VO and throughout the whole temperature range (100 K $\lesssim$ T $\lesssim$ 6000 K) for VO$_2$. Using the JANAF tables and a solar elemental composition, VO$_2$ is the most abundant vanadium-bearing species for temperatures below $\sim$1500 K. Between $\sim$1500 K and $\sim$2000 K VO becomes the most abundant V-containing species and for higher temperatures it is atomic vanadium. With the updated thermochemical data the behavior of VO and VO$_2$ is flipped for temperatures between $\sim$300 K and $\sim$1500 K. Due to the updated higher Gibbs free energies the fractional abundance of VO$_2$ has decreased throughout the entire temperature range. This behavior is consistent with observations. VO and atomic vanadium have both been detected in hot Jupiters \citep{pelletier_vanadium_2023} whereas the presence of VO$_2$ has been predicted based on previous equilibrium chemistry calculations \citep{hoeijmakers_hot_2020} but not yet detected. Owing to element conservation the abundances of other V-bearing species also are affected. This effect can be seen in the abundances of e.g. V and VN.\\
\begin{figure}
\centering
\includegraphics[width=9.6cm]{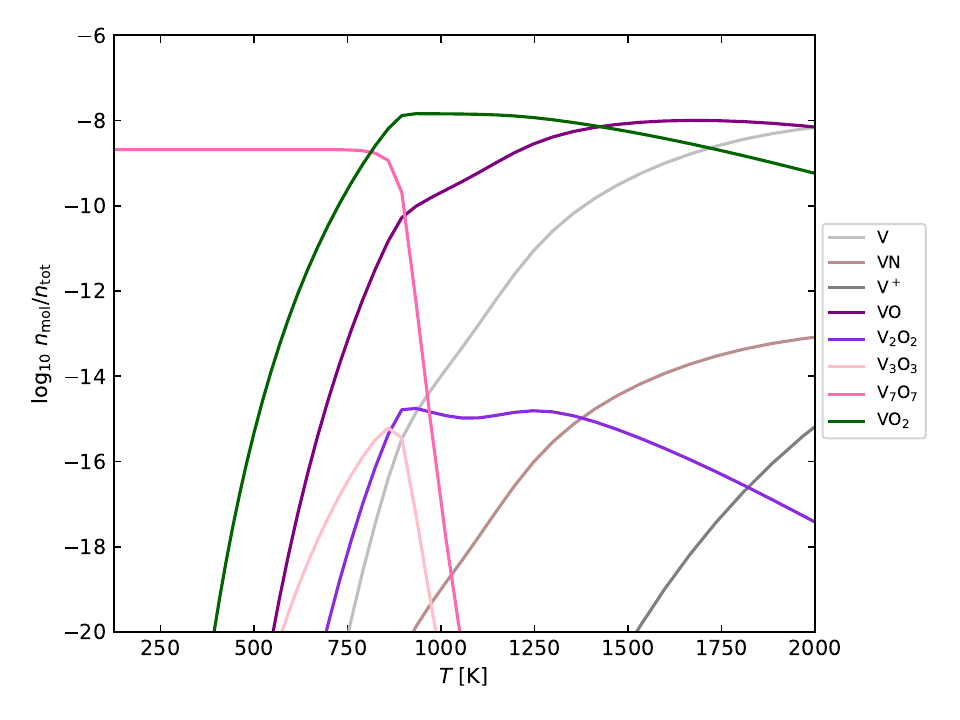}
\includegraphics[width=9.6cm]{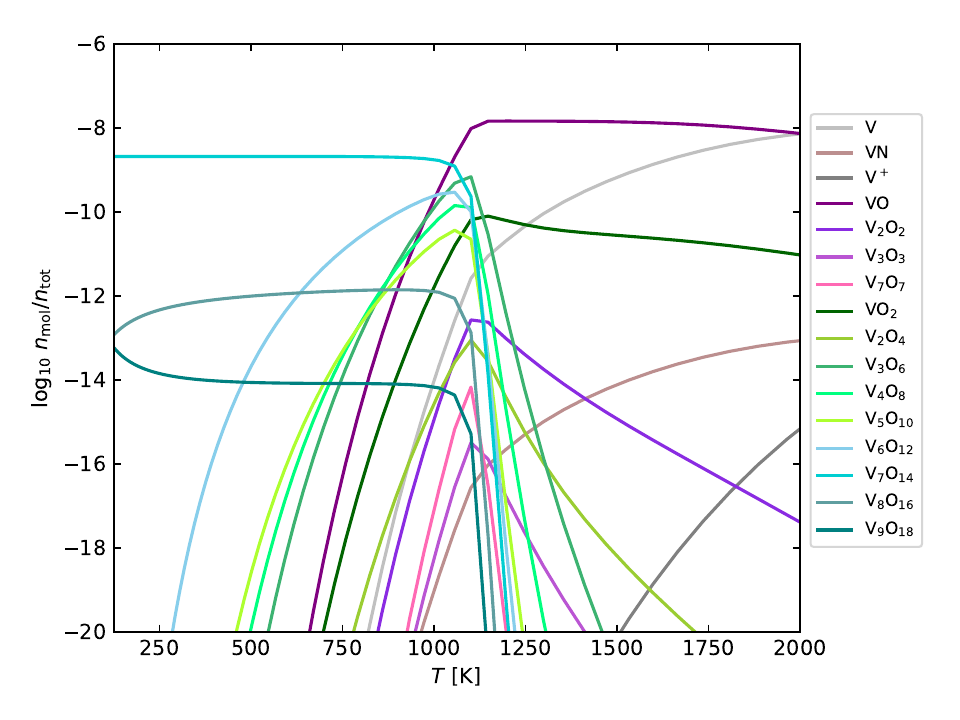}
\includegraphics[width=9.6cm]{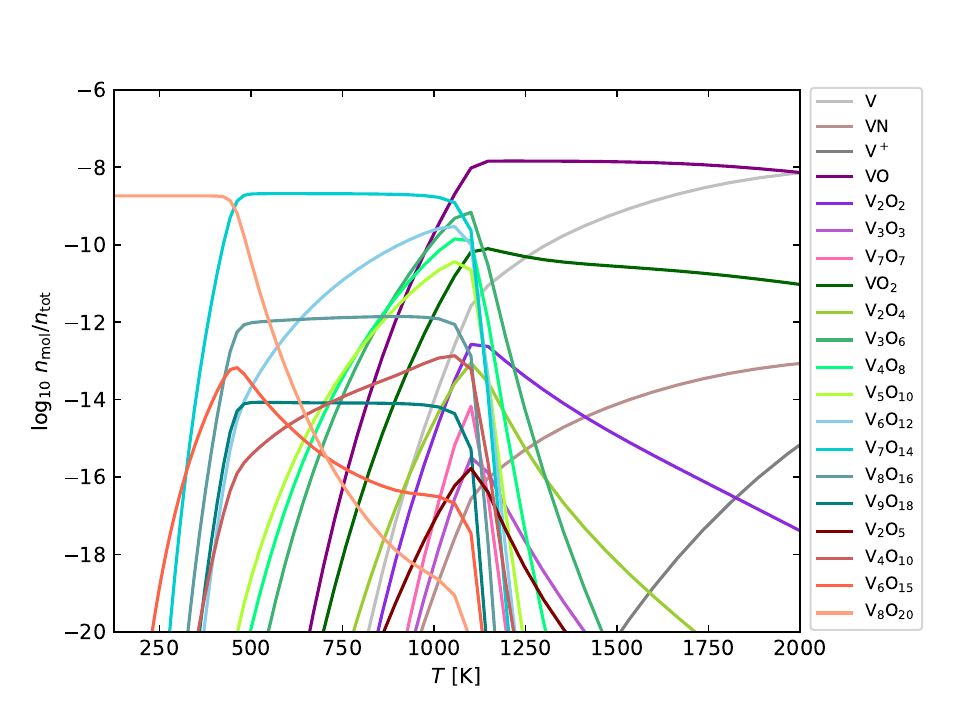}
  \caption{Abundances of all the vanadium-bearing species already present in \texttt{GGchem} and the clusters obtained from our DFT calculations as a function of temperature. \textbf{Top}: all (VO)\textsubscript{N} clusters were included, \textbf{middle}: all (VO)\textsubscript{N}  and  (VO$_2$)\textsubscript{N}  clusters were included, \textbf{bottom}:  all (VO)\textsubscript{N}, (VO$_2$)\textsubscript{N} and (V$_2$O$_5$)\textsubscript{N} clusters were included. The calculations were again ran at $P=$ 1 bar and with solar elemental abundances from \citet{asplund_chemical_2009}.}
\label{fig:GGchem_1D_results}
\end{figure}
Since we aim to model CCN formation, it is necessary to study clusters with various V:O stoichiometries and larger sizes that go beyond the respective monomers and species currently included in \texttt{GGchem}.  
In Fig. \ref{fig:GGchem_1D_results} the chemical equilibrium abundances of the vanadium-bearing species are shown for a pressure of 1 bar and a gas temperature range between 100 and 2000 K. In the top panel the abundances of vanadium-bearing molecules including the (VO)\textsubscript{N} stoichiometric cluster family are displayed. In the middle panel, the (VO$_2$)\textsubscript{N} clusters are additionally included. The lower panel of Fig. \ref{fig:GGchem_1D_results} shows the abundances of V-containing species including all considered cluster families, i.e. (VO)\textsubscript{N}, VO$_2$)\textsubscript{N}, and V$_2$O$_5$)\textsubscript{N}. In a stepwise manner each cluster was added individually in order of increasing size. We found that each newly added vanadium monoxide (VO)\textsubscript{N} cluster (Fig. \ref{fig:GGchem_1D_results} (top)) becomes the most abundant species below $\sim$1000 K up to a size of N=7. For N$>$7 this is not the case anymore. Such behaviour may be due to the higher stability of (VO)$_7$ shown in Fig. \ref{fig:binding_energy} (top) and to the fact that the 1:1 ratio is not favored for bigger sizes. For the (VO$_2$)\textsubscript{N} GM candidates  (Fig. \ref{fig:GGchem_1D_results} (middle)) the situation is similar as the heptamer (N=7) also corresponds to the most abundant vanadium oxide cluster. The (V$_2$O$_5$)\textsubscript{N} N=1-4 cluster subset (Fig. \ref{fig:GGchem_1D_results} (bottom)) becomes the most abundant V-bearing species for temperatures below $\sim$500 K, leaving (VO$_2$)$_7$ as the most abundant species between $\sim$500 K and $\sim$1000 K, and the vanadium monoxide monomer and atomic vanadium as predominant species for higher temperatures. Since the latter case (Fig. \ref{fig:GGchem_1D_results} (bottom)) corresponds to the most extensive set of V-bearing species,  we consider the resulting equilibrium abundances as most realistic. We note that vanadium monoxide is the only monomer more abundant than the clusters of its stoichiometric family, indicating 
it may remain detectable after the nucleation process has taken place. Contrary, the spectral features of the VO$_2$ and V$_2$O$_5$ monomers may be hidden by the spectral features of the most stable clusters in their respective stoichiometric families.

\begin{figure}
\centering
\includegraphics[width=9cm]{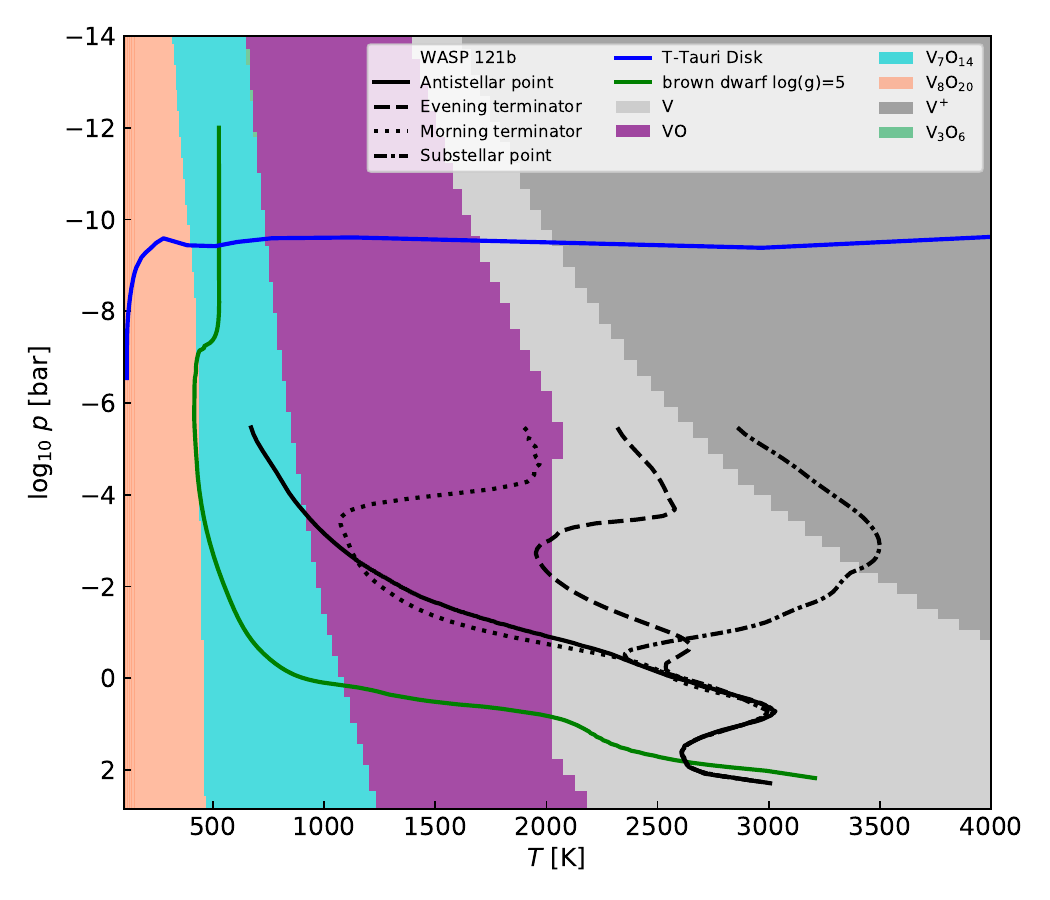}
  \caption{Most abundant vanadium-bearing molecules in chemical equilibrium for a grid of gas pressures and temperatures. The model assumes solar elemental abundances from \cite{asplund_chemical_2009} as initial conditions. (T$_{gas}$, p$_{gas}$) profiles of the hot Jupiter WASP-121b (\cite{helling_cloud_2021}, \cite{parmentier_thermal_2018}, \cite{mansfield_hstwfc3_2018}), a brown dwarf (\cite{dehn_self-consistent_2007}, \cite{helling_dust_2008}, \cite{witte_dust_2009}, \cite{witte_dust_2011}) and a T-Tauri protoplanetary disk \citep{kanwar_notitle_2023} are over plotted.}
\label{fig:2D_TPprofiles}
\end{figure}
\noindent In order to account for the conditions of cloud formation in substellar atmospheres we have applied \texttt{GGchem} for a grid of gas pressures and gas temperatures in the ranges (700, 1$\cdot$ 10$^{-14}$) bar and (100, 4000) K respectively. In Fig. \ref{fig:2D_TPprofiles} we plot the pressure-temperature (T$_{gas}$, p$_{gas}$) profiles of the Hot Jupiter WASP-121 b \citep{helling_cloud_2021, parmentier_thermal_2018, mansfield_hstwfc3_2018}) in black, a brown dwarf with  log(\textit{g})=5 \citep{dehn_self-consistent_2007,helling_dust_2008, witte_dust_2009,witte_dust_2011} in green and a T-Tauri protoplanetary disk \citep{kanwar_notitle_2023} in blue over the pressure temperature grid results.\\
WASP-121 b is an ultra-hot Jupiter (UHJ), with a global temperature of approximately 2700 K \citep{evans_detection_2016} and therefore its evening terminator and substellar point are too hot for the clusters to be abundant \citep{parmentier_thermal_2018} and their chemistry to be relevant, which is in agreement with our equilibrium calculations. We believe the role of clusters in the formation of 
CCN will become more relevant for cooler objects, such as the brown dwarf in Fig. \ref{fig:2D_TPprofiles}, the cooler regions to the midplane of protoplanetary disks (the (T$_{gas}$, p$_{gas}$) profile shown in Fig. \ref{fig:2D_TPprofiles} corresponds to a cut from the midplane outwards at a distance of 1 AU), as well as cooler exoplanets such as colder hot jupiters or warm saturns. We have chosen WASP-121 b as the Hot Jupiter because, together with TiO, VO has been considered as a possible driver for the temperature inversion in the atmosphere, even though their detection and their role as main drivers of the thermal inversion remains disputed \citep{evans_detection_2016, 
ouyang_detection_2023, mikal-evans_confirmation_2020, mikal-evans_emission_2019, merritt_non-detection_2020}.
The related observed transmission spectra are constrained to optical and near-infrared wavelengths and do not probe transitions in VO with lower energies. Our findings predict VO as the most abundant vanadium-bearing species for a considerable extension of the atmosphere at the antistellar point (solid black line) and at the morning terminator (dotted black line) using chemical equilibrium and assuming that the atmosphere is cloud free. 
We therefore conclude that it is very likely that vanadium monoxide is present in the atmosphere of WASP-121 b.\\ 

\section{Astrophysical Relevance}
\label{sect:astro_rel}

In this section we asses the viability of vanadium oxides as condensation seed candidates, as well as their possible observability. We apply different nucleation theories to determine the nucleation rate of the three considered vanadia stoichiometric families taking into account the thermochemical data available for each of them. All the nucleation rates were calculated along the (T$_{gas}$, p$_{gas}$) profile of the brown dwarf with log(g)=5 from Fig. \ref{fig:2D_TPprofiles} (green line). The gas-phase composition was calculated with the equilibrium chemistry code \texttt{GGchem} assuming that all the monomers (VO, VO$_2$ and V$_2$O$_5$) and the additional vanadium bearing species from Fig. \ref{fig:comparison_GGchem} are present in the gas-phase. Futhermore, we have extracted the vibrational frequencies from our DFT calculations to generate vibrational spectra and determine the wavelength range at which the most intense emission peaks may appear. We find that all the emission peaks are within the JWST-MIRI detection range.  

\begin{figure}
\centering
\includegraphics[width=9cm]{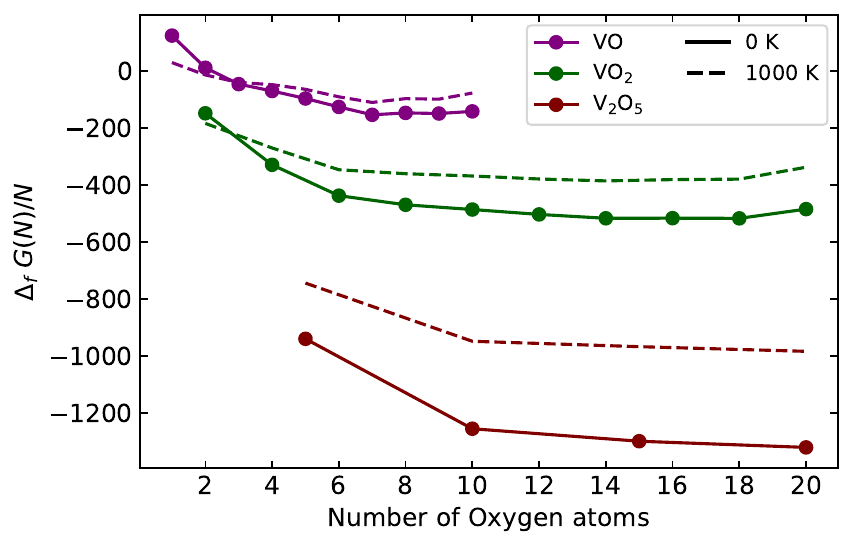}
  \caption{Comparison of the Gibbs free energies of formation normalized to the cluster size ($\Delta_f$G(N)/N) for the clusters (VO)\textsubscript{N} and (VO$_2$)\textsubscript{N} with N=1-10, (V$_2$O$_5$)\textsubscript{N} with N=1-4 as a function of degree of oxidation for T=0 K (\textbf{full lines}) and T=1000 K (\textbf{dashed lines}).}
\label{fig:comparison_stab_stoichiometries}
\end{figure}

\subsection{Nucleation rates of vanadium oxides}
\label{sec:results_nuc_rates}
\begin{figure}
\center
\includegraphics[width=9cm]{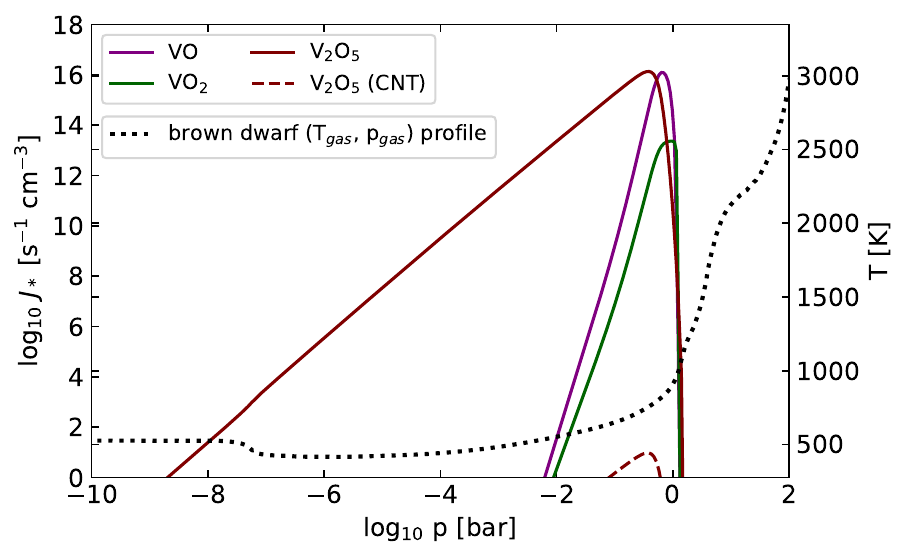}
  \caption{\textbf{Full lines}: Non-classical nucleation rate (Eq. \ref{eq:rate_nonclassical]}) for VO, VO$_2$ and V$_2$O$_5$ as a function of the pressure. \textbf{Dashed line}: classical nucleation rate (Eq. \ref{eq:nucleationrate_CNT}) for V$_2$O$_5$. The rates have been calculated following the (T$_{gas}$, p$_{gas}$) profile of a brown dwarf with log(g)=5 (black dotted line). }
\label{fig:nonclassical_rates}
\end{figure}

The difference in the free energies between the different vanadium oxide stoichiometries, i.e. oxidation states, (Fig. \ref{fig:comparison_stab_stoichiometries}) indicates that the formation pathways from atomic vanadium to a CCN consisting of hundreds to thousands of atoms is likely not through monomeric homomolecular addition. 
We will provide a first step towards understanding vanadium nucleation by applying different nucleation theories for the three stoichiometric families considered in this study. 
Nucleation theories such as CNT and MCNT assume monomeric homomolecular addition and 
therefore give an incomplete image of vanadia nucleation, but they represent a first step towards a fully 
kinetic approach. 
Since the thermodynamically most stable vanadium oxide stoichiometry in the solid phase is V$_2$O$_5$ we applied classical nucleation theory to this cluster family and calculate its surface tension and nucleation rate. We applied non-classical nucleation theory, which relies 
on the actual energies of the clusters, to (VO)\textsubscript{N}, (VO$_2$)\textsubscript{N} and (V$_2$O$_5$)\textsubscript{N}. \\
We have followed the approach from \cite{lee_dust_2014} and \cite{sindel_revisiting_2022} to calculate the surface tension and the CNT nucleation rate. A detailed explanation, as well as the surface tension fit, can be found in App. \ref{appendix_surfacetension}. Since V$_2$O$_5$ 
undergoes a phase transition from crystalline to liquid at T= 943 K we have fitted the surface tension for the range T=0 K to T=900 K:
\begin{equation}
    \sigma_{\infty}=110.14-0.0484 \cdot \text{T}
\end{equation}
Given the range of the surface tension, we are limited to calculate the CNT nucleation rate for objects cooler than 900 K. For comparison purposes we have chosen to calculated all the nucleation rates for the brown dwarf (T$_{gas}$, p$_{gas}$) profile from Fig. \ref{fig:2D_TPprofiles} and the results can be seen on Fig. \ref{fig:nonclassical_rates}.

\noindent (VO)\textsubscript{N} and (VO$_2$)\textsubscript{N} are not the preferred oxidation state in the bulk phase and therefore the available solid phase data is less reliable. For this reason we have decided to calculate their nucleation rates applying only non-classical nucleation theory. We have applied the Becker-D\"{o}ring method \citep{gail_growth_2013,lee_dust_2014, sindel_revisiting_2022} to obtain the summation:
\begin{equation}
    J_*^{-1}(\text{T})=\sum_{\text{N}=1}^{\text{N}_{\text{max}}} \left( \frac{\tau_{gr}(r_i,\text{N},\text{T})}{ \mathring{f}(\text{N})}    \right)
    \label{eq:rate_nonclassical]}
\end{equation}
where $\tau_{gr}$ corresponds to Eq. \ref{eq:tau_growth} and the equilibrium number density of a cluster of size N can be calculated from the partial pressures
\begin{equation}
    \mathring{f}(\text{N})=\frac{\mathring{p}(\text{N})}{k\text{T}}
\end{equation}
that we obtain from applying the law of mass action to a cluster of size N
\begin{equation}
    \mathring{p}(\textrm{N})=p^{\plimsoll} \left( \frac{\mathring{p}(1)}{p^{\plimsoll}}\right)^{\text{N}} \textrm{exp} \left(-\frac{\Delta_f^{\plimsoll}G(\text{N})-\text{N}\Delta_f^{\plimsoll}G(1)}{R\text{T}} \right)
\end{equation}
where the $\Delta_f^{\plimsoll}$G(N) are derived from DFT calculations and the partial pressure of the monomer ($\mathring{p}$ (1) [dyn cm$^{-2}$]) from our equilibrium data assuming local thermodynamic equilibrium (LTE). The results are summarized in Fig. \ref{fig:nonclassical_rates}.

\subsection{Vibrational Synthetic Spectra}

\begin{figure}
\centering
\includegraphics[width=0.5\textwidth]
{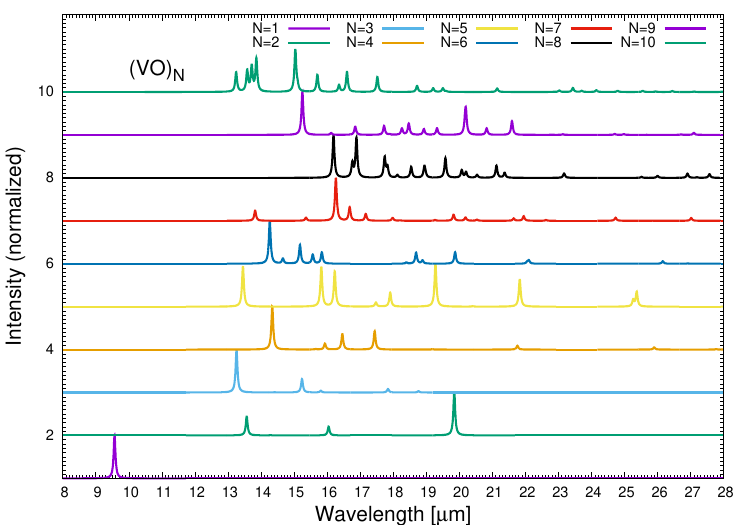}
\includegraphics[width=0.5\textwidth]
{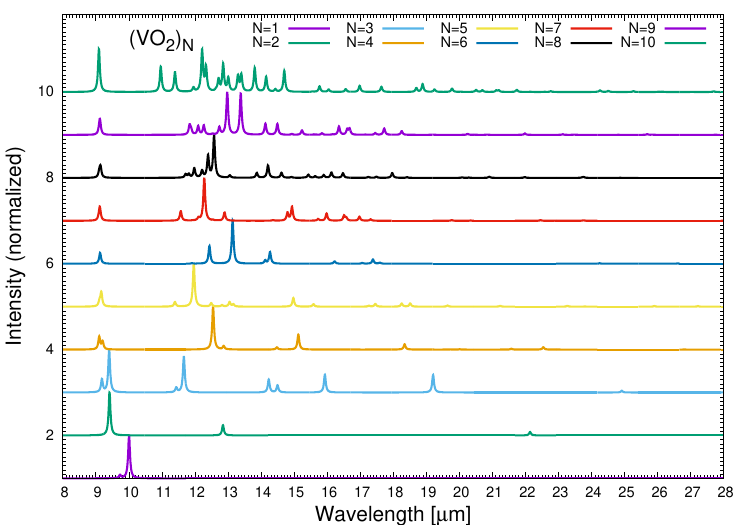}
\includegraphics[width=0.5\textwidth]
{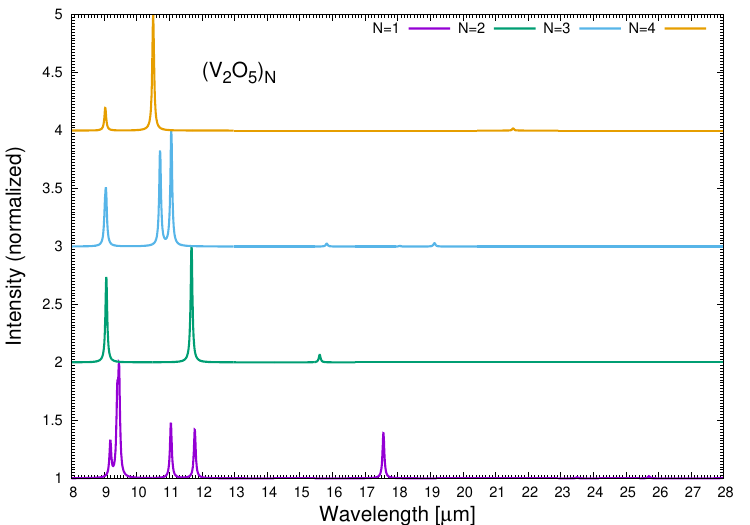}
\caption{Normalized intensities of the vibration modes of the GM cluster candidates as a function of the wavelength (in $\mu$m) for \textbf{top panel:} (VO)\textsubscript{N}, N=1-10, \textbf{middle panel:} (VO$_2$)\textsubscript{N}, N=1-10, and \textbf{bottom panel:} (V$_2$O$_5$)\textsubscript{N}, N=1-4.
\label{spectra}}
\end{figure}

The VO molecule shows strong transitions between electronically excited A$^4\Pi$, B$^4\Pi$, and C$^4\Sigma^-$ states and the ground state C$^4\Sigma^-$ that are located at 1055 nm, 790 nm and 574 nm \citep{karlsson_lifetime_1997}.
In our current study electronically excited states are not investigated, but internal vibrations and rotations are considered.  
According to our calculations VO exhibits an intense vibrational mode at 9.569 $\mu$m and has a rotational constant of 16646.5 MHz. Related internal transitions of VO, if present, are potentially detectable in planetary atmospheres with state-of-the-art observing facilities like JWST.\\
The vibrational frequencies of the GM cluster candidates, that are required to calculate the partition functions and thermodynamic potentials, also serve us to generate synthetic spectra. In Fig. \ref{spectra} we show these spectra for the different cluster families and 
sizes considered in this study. Here, we assume Lorentzian line profiles with a full width at half maximum (FWHM) of 0.033 $\mu$m corresponding to a typical ALMA setup resolution. The line intensities are normalized to its maximum value for each cluster size. For (VO)\textsubscript{N}, N=2-10, the most intense vibration modes occur in a wavelength range between 13 $\mu$m and 20 $\mu$m, whereas the VO monomer represents an exception. Overall, we do not find common spectral features for the different (VO)\textsubscript{N} sizes.\\
The (VO$_2$)\textsubscript{N}, N=1-3, GM candidates show their maximum emission below $\sim$10 $\mu$m, while the larger 
clusters with N=4$-$10 emit most intensely between 12.0 $\mu$m and 13.2 $\mu$m. These larger clusters all show a common feature at around 9.1 $\mu$m that can be attributed to the stretching modes of the terminal, singly coordinated oxygen atom in these structures.\\
The four (V$_2$O$_5$)\textsubscript{N} clusters show only a few distinct peaks in the infrared spectra, as their symmetric lowest-energy 
structures have several vibrational modes with identical frequencies. Their maximum emission lies in the range of 9 to 
12 $\mu$m. The crystalline V$_2$O$_5$ spectra from \citet{abello_vibrational_1983} shows prominent emission at $\sim$ 9.7 microns, similarly to our GM candidates. We note that the bulk spectral lines are broad and blurry in comparison with the discrete and narrow emission lines of the clusters. Although an identification of a specific cluster in a real observed spectra might be challenging, the different cluster stoichiometric families show distinct spectral characteristics and could be distinguished from each other.\\
We also note that a number of approximations are used here, including the RRHO approximation at T=0 K, a uniform line broadening, and a background noise that is independent of the planet or host star. We aim to address the spectral properties of the clusters in more detail in future kinetic study.

\section{Conclusions}
\label{sect:conclusions}
The nucleation process is an important part of cloud formation in substellar atmospheres and dust formation in several other astrophysical environments. Presently, the nature of the nucleation seeds, as well as their formation process, is not well understood. \\
This paper investigates the nucleation properties of vanadium oxides in substellar atmospheres, since both atomic vanadium and gaseous VO have been detected in substellar objects but have not previously been considered as nucleation species.\\
We have used a hierarchical optimization approach to obtain hitherto unreported vanadium oxide cluster structures for the stoichiometries: (VO)\textsubscript{N} and (VO$_2$)\textsubscript{N} with N=1-10 and (V$_2$O$_5$)\textsubscript{N} with N=1-4. Density functional theory at the B3LYP/cc-pVTZ level has been applied to obtain the final energies, thermochemical potentials and vibrational frequencies.\\
The impact of revised and more accurate thermochemical potentials is studied in chemical equilibrium with the code \texttt{GGchem}. Our 
results suggest that even small changes in the Gibbs free energies of formation of the monomers can result in large changes on the abundances of vanadium bearing species at low temperatures (below $\sim$1000 K). The consideration of larger cluster species also vastly modifies the equilibrium abundances, giving more realistic results and indicating thermodynamically viable cluster and dust formation routes. The updated abundances provide insights on which species could be observed by facilities such as the James Webb Space Telescope.\\
Vanadium oxide clusters become predominant at temperatures around/below 1000 K which makes them relevant for atmospheres of hot Jupiters as well as other astrophysical objects like brown dwarfs, AGB stars and protoplanetary disks. We have calculated the vibrational spectra of each cluster and found major emission peaks at wavelengths between 8 and 28 microns, which are within the JWST-MIRI detection range. We have also calculated nucleation rates for a brown dwarf with log(g)=5. Due to the difference in the free energies of formation of the different vanadium stoichiometries, the nucleation process most likely does not follow monomeric homomolecular addition and therefore current theories such as CNT and MCNT do not provide an accurate description of the process. Since V$_2$O$_5$ is the most stable oxidation state solid bulk phase data is available and we were able to apply CNT to obtain a nucleation rate via the surface tension. The CNT nucleation rate is approximately 15 orders of magnitude lower than the non-classical nucleation rate for V$_2$O$_5$. We were only able to obtain non-classical nucleation rates for VO and VO$_2$ and those are comparable to the V$_2$O$_5$ non-classical rate. Our results suggest that, even though CNT is an useful tool to calculate nucleation rates when the thermochemical data for clusters is not available, taking such data into account can increase the nucleation rate of a given species by several orders of magnitude. A full chemical-kinetic approach is needed to explore the efficiency of vanadium oxides as heterogeneous nucleation candidates and compare them with the nucleation species used in other studies.

\begin{acknowledgements}
    H.L.M, L.D and Ch.H. acknowledge funding from the European Union H2020-MSCA-ITN-2019 under grant agreement no. 860470 (CHAMELEON). L.D acknowledges support from the ERC consolidator grant 646758 AEROSOL and the KU Leuven IDN/19/028 grant Escher. D.G. was funded by the project grant ‘The Origin and Fate of Dust in Our Universe’ (research grant KAW 2020.0081) from the Knut and Alice Wallenberg Foundation. J.P.S. acknowledges a St Leonard’s Global Doctoral Scholarship from the University of St Andrews and funding from the Austrian Academy of Science.The computational results have been obtained using the Vienna Scientific Cluster and the St. Andrews HPC KENNEDY cluster. The computations involved the Swedish National Infrastructure for Computing (SNIC) at Chalmers Centre for Computational 
Science and Engineering (C3SE) partially funded by the Swedish Research Council through grant no. 2018-05973.
\end{acknowledgements}
\bibliographystyle{aa} 
\bibliography{references} 

\begin{appendix} 
\onecolumn
\section{GGchem fit parameters}
\label{appendix_GGchem}

\begin{table*}[h]
\begin{center}
\caption{Parameters $a_0$, $a_1$, $b_0$, $b_1$ and $b_2$ implemented in \texttt{GGchem} to compute the global minimization.}
\begin{tabular}{cccccc}
\hline
Formula          & $a_0$    & $a_1$     & $b_0$     & $b_1$    & $b_2$     \\ \hline
VO               & 7.48E+04 & -7.73E+00 & 3.74E+01  & 4.42E-03 & -3.82E-07 \\
V$_2$O$_2$       & 1.81E+05 & -3.26E+00 & -2.34E+01 & 1.07E-03 & -8.18E-08 \\
V$_3$O$_3$       & 2.92E+05 & -3.85E+00 & -5.35E+01 & 1.59E-03 & -1.22E-07 \\
V$_4$O$_4$       & 4.01E+05 & -4.33E+00 & -8.51E+01 & 2.09E-03 & -1.61E-07 \\
V$_5$O$_5$       & 5.18E+05 & -4.78E+00 & -1.17E+02 & 2.56E-03 & -1.98E-07 \\
V$_6$O$_6$       & 6.43E+05 & -5.40E+00 & -1.45E+02 & 3.10E-03 & -2.39E-07 \\
V$_7$O$_7$       & 7.73E+05 & -6.10E+00 & -1.77E+02 & 3.67E-03 & -2.82E-07 \\
V$_8$O$_8$       & 8.77E+05 & -6.64E+00 & -2.11E+02 & 4.18E-03 & -3.22E-07 \\
V$_9$O$_9$       & 9.89E+05 & -7.02E+00 & -2.41E+02 & 4.63E-03 & -3.58E-07 \\
V$_{10}$O$_{10}$ & 1.09E+06 & -9.28E+00 & -2.76E+02 & 5.79E-03 & -4.35E-07 \\ \hline
VO$_2$           & 1.39E+05 & -2.75E+00 & -1.00E+01 & 7.54E-04 & -5.35E-08 \\
V$_2$O$_4$       & 3.23E+05 & -4.00E+00 & -5.40E+01 & 1.72E-03 & -1.20E-07 \\
V$_3$O$_6$       & 5.24E+05 & -5.22E+00 & -9.84E+01 & 2.67E-03 & -1.85E-07 \\
V$_4$O$_8$       & 7.14E+05 & -6.42E+00 & -1.44E+02 & 3.63E-03 & -2.51E-07 \\
V$_5$O$_{10}$      & 9.02E+05 & -7.53E+00 & -1.88E+02 & 4.54E-03 & -3.14E-07 \\
V$_6$O$_{12}$      & 1.10E+06 & -8.83E+00 & -2.33E+02 & 5.53E-03 & -3.82E-07 \\
V$_7$O$_{14}$      & 1.29E+06 & -9.65E+00 & -2.82E+02 & 6.33E-03 & -4.39E-07 \\
V$_8$O$_{16}$      & 1.47E+06 & -1.09E+01 & -3.27E+02 & 7.28E-03 & -5.05E-07 \\
V$_9$O$_{18}$      & 1.66E+06 & -1.17E+01 & -3.74E+02 & 8.12E-03 & -5.64E-07 \\
V$_{10}$O$_{20}$     & 1.80E+06 & -1.31E+01 & -4.27E+02 & 9.13E-03 & -6.34E-07 \\ \hline
V$_2$O$_5$       & 3.87E+05 & -4.62E+00 & -6.69E+01 & 2.13E-03 & -1.43E-07 \\
V$_4$O$_{10}$      & 8.50E+05 & -7.38E+00 & -1.74E+02 & 4.34E-03 & -2.93E-07 \\
V$_6$O$_{15}$      & 1.29E+06 & -1.03E+01 & -2.76E+02 & 6.59E-03 & -4.43E-07 \\
V$_8$O$_{20}$      & 1.73E+06 & -1.30E+01 & -3.75E+02 & 8.73E-03 & -5.88E-07 \\ \hline
\end{tabular}
\end{center}
\label{tb:DelG_coef}
\end{table*}

\twocolumn
\section{Surface tension calculations}
\label{appendix_surfacetension}

In classical nucleation theory the stationary nucleation rate is calculated according to
\begin{equation}
    J_{*}^{C}=\frac{f^{\circ}(1)}{\tau_{gr}(r_i,\text{N}_*,\textrm{T} )}Z(\text{N}_*) \textrm{exp} \left( (\text{N}_*-1) \textrm{ln}\ S(\textrm{T} ) -\frac{\Delta G(\text{N}_*)}{R\textrm{T} } \right)
    \label{eq:nucleationrate_CNT}
\end{equation}
where $S(\textrm{T})$ is the supersaturation ratio, $f^{\circ}$(1) is the monomer number density, $\tau_{gr}$ is the growth time scale, N$_*$ is the critical cluster size and Z(N$_*$) is the Zeldovich factor. We calculated the growth time scale according to
\begin{equation}
    \tau_{gr}^{-1}= A(\text{N})\alpha (\text{N}) \nu_{rel}n_{f}
    \label{eq:tau_growth}
\end{equation}
where A(N)=4$\pi a_0^2N^{2/3}$ is the effective cross section of a spherical (V\textsubscript{x}O\textsubscript{y})\textsubscript{N} cluster, $\alpha$ is the sticking coefficient (assumed to be 1), $n_f$ is the monomer number density and $\nu_{rel}$ is the relative velocity defined as:
\begin{equation}
    \nu_{rel}=\sqrt{\frac{k\textrm{T}}{2\pi m_x}}
\end{equation}
with $m_x$ the mass of the monomer and $k$ the Boltzmann constant. The Zeldovich factor accounts for the contribution from Brownian motion to the nucleation process and it can be calculated as:
\begin{equation}
    Z(\text{N}_*)=\sqrt{\frac{\theta_{\infty}}{9\pi(\text{N}_*-1)^{4/3}}}
\end{equation}
with $\theta_{\infty}$ defined as
\begin{equation}
    \theta_{\infty}=\frac{4\pi a_0^2 \sigma_{\infty}}{k_b\textrm{T} }
    \label{eq:theta_infty}
\end{equation}
in classical nucleation theory. a$_0$ is the theoretical monomer radius which can be calculated from the bulk properties
\begin{equation}
    a_0=\left( \frac{3M_{V_xO_y}}{4\pi\rho_{V_xO_y}}  \right)^{1/3}
\end{equation}
and $\sigma_{\infty}$ is the surface tension, also dependent on the properties of the bulk solid. The final term of equation Eq. \ref{eq:nucleationrate_CNT} corresponds to the Gibbs free energies of formation of the critical cluster and we will use the values from our DFT calculations. 
Since the most stable vanadium oxide stoichiometry in the solid phase is V$_2$O$_5$ we will only apply classical nucleation theory to the clusters from this stoichiometry. 
In order to eventually obtain a nucleation rate we need to first compute the surface tension of the bulk solid $\sigma_{\infty}$ which is the quantity used to measure the impact of the Gibbs free energies of formation in the nucleation process. We will follow the approach of \cite{jeong_electronic_2000} as done in \cite{sindel_revisiting_2022} and \cite{lee_dust_2014}, where the nucleation is linked to the Gibbs free energies through:
\begin{figure}
\center
\includegraphics[width=9cm]{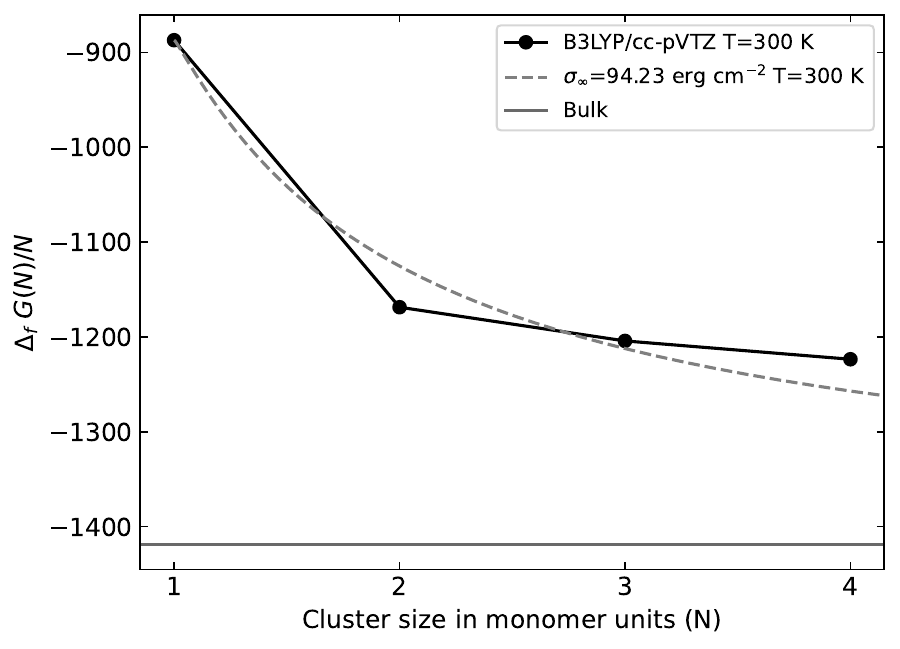}
  \caption{Gibbs free energies of formation for the clusters (V$_2$O$_5$)\textsubscript{N} with N=1-4 at T=1000. We have calculate a fit for $\sigma_{\infty}$ according to Eq. \ref{eq:surface_tension}}
\label{fig:surfacetension_V2O5}
\end{figure}
\begin{small}
\begin{equation}
    \frac{\Delta_f G^{\circ}(\text{N})}{\text{N}}= \left( \theta_{\infty}R\textrm{T}\frac{\text{N}-1}{(\text{N}-1)^{1/3}+\text{N}_{f}^{1/3}} + \Delta_f G^{\circ}(1) + (\text{N}-1)\Delta_{f} G^{\circ}(s)    \right) \cdot \text{N}^{-1}
    \label{eq:surface_tension}
\end{equation}
\end{small}and $\Delta_f$G$^{\circ}$(N) is the Gibbs free energy of formation of cluster size N,  $\Delta_f$G$^{\circ}$(1) is the Gibbs free energy of formation of the monomer,  $\Delta_f$G$_1^{\circ}$(s) is the Gibbs free energy of the bulk phase, N$_f$ is a fitting factor that will be set to zero and $\theta_{\infty}$ as been defined in Eq.\ref{eq:theta_infty}. We have calculated $a_0$ for the V$_2$O$_5$ monomer:
\begin{equation}
    a_0=\left( \frac{3M_{V_2O_5}}{4\pi\rho_{V_2O_5}}  \right)^{1/3} \simeq 2.782 \times 10^{-10} \ m
\end{equation}
This value approximately corresponds to half of the length between the most distant atomic centers in the V$_2$O$_5$ monomer, i.e $\frac{5.191\AA{}}{2}$. We note that the V$_2$O$_5$ cluster is not spherical (Fig. \ref{fig:geometries}(u)). We have obtained a surface tension of 94.23 erg$\cdot$cm$^{-2}$ at T=300 K. The value of $\sigma_{\infty}$ is dependent on pressure. For the range T=0 K to T= 900 K we obtain the following fit:
\begin{equation}
    \sigma_{\infty}=110.14-0.0484 \cdot \textrm{T} 
\end{equation}
The fitting range was chosen taking into account that V$_2$O$_5$ transitions from liquid to solid at T= 943 K. The surface tension has been used to calculate the nucleation rate along the (T$_{gas}$, p$_{gas}$) profile of the brown dwarf from Fig. \ref{fig:2D_TPprofiles} (green line) and the results are discussed in Sect. \ref{sec:results_nuc_rates}.

\section{Rotational Constants}
\label{appendix_RotC}

\begin{table}[]
\begin{tabular}{cccc}
{N}                & \multicolumn{3}{c}{{Rotational Constants (GHz)}} \\ \hline
{(VO)\textsubscript{N}}         & {}         & {}        & {}        \\ \hline
{1}                & {16.646}   & {}        & {}        \\
{2}                & {9.199}    & {2.972}   & {2.246}   \\
{3}                & {1.971}    & {1.517}   & {0.857}   \\
{4}                & {1.659}    & {0.660}   & {0.517}   \\
{5}                & {0.850}    & {0.558}   & {0.463}   \\
{6}                & {0.410}    & {0.367}   & {0.340}   \\
{7}                & {0.360}    & {0.282}   & {0.225}   \\
{8}                & {0.342}    & {0.212}   & {0.204}   \\
{9}                & {0.293}    & {0.138}   & {0.125}   \\
{10}               & {0.268}    & {0.166}   & {0.139}   \\ \hline
{(VO$_2$)\textsubscript{N}}     & {}         & {}        & {}        \\ \hline
{1}                & {35.512}   & {8.434}   & {6.815}   \\
{2}                & {7.630}    & {1.276}   & {1.165}   \\
{3}                & {1.303}    & {0.908}   & {0.613}   \\
{4}                & {0.731}    & {0.446}   & {0.429}   \\
{5}                & {0.646}    & {0.235}   & {0.187}   \\
{6}                & {0.522}    & {0.136}   & {0.120}   \\
{7}                & {0.310}    & {0.133}   & {0.127}   \\
{8}                & {0.239}    & {0.101}   & {0.092}   \\
{9}                & {0.218}    & {0.069}   & {0.068}   \\
{10}               & {0.127}    & {0.097}   & {0.090}   \\ \hline
{(V$_2$O$_5$)\textsubscript{N}} & {}         & {}        & {}        \\ \hline
{1}                & {4.250}    & {1.035}   & {0.995}   \\
{2}                & {0.404}    & {0.404}   & {0.404}   \\
{3}                & {0.201}    & {0.158}   & {0.158}   \\
{4}                & {0.097}    & {0.097}   & {0.097}  
\end{tabular}
\end{table}

\end{appendix}

\end{document}